\documentclass[showpacs,preprintnumbers,amsmath,amssymb]{revtex4}

\usepackage{epsfig}
\usepackage{graphicx}
\usepackage{dcolumn}
\usepackage{bm}

\begin{document}


\title{General spectral function expressions of a 1D correlated model}
\author{J. M. P. Carmelo}
\affiliation{Department of Physics, Massachusetts Institute of Technology, Cambridge,
Massachusetts 02139-4307, USA} \affiliation{GCEP - Center of Physics, University of
Minho, Campus Gualtar, P-4710-057 Braga, Portugal}
\author{K. Penc}
\affiliation{Research Institute for Solid State Physics and Optics, H-1525 Budapest,
P.O.B. 49, Hungary}
\date{6 August 2005}

\begin{abstract}
We introduce a method that allows the evaluation of general expressions for the spectral
functions of the one-dimensional Hubbard model for {\it all} values of the on-site
electronic repulsion $U$. The spectral weights are expressed in terms of pseudofermion
operators such that the spectral functions can be written as a convolution of
pseudofermion dynamical correlation functions. Our results are valid for {\it all} finite
energy and momentum values and are used elsewhere in the study of the unusual
finite-energy properties of quasi-one-dimensional compounds and the new quantum systems
of ultra-cold fermionic atoms on an optical lattice.
\end{abstract}


\maketitle
\section{INTRODUCTION}
\label{SecI}

The main goal of this paper is to provide a general method for the evaluation of matrix
elements of one-, two-electron, or $\cal{N}$-electron operators such that $\cal{N}$ is
finite, between the ground state and excited energy eigenstates of the one-dimensional
(1D) Hubbard model. Our results correspond to an important part of the derivation of the
one-electron and two-electron spectral-weight distributions used in Refs.
\cite{super,spectral1} in the study of the phase diagram and unusual one-electron
spectral properties of quasi-1D compounds. Indeed, the matrix-element and general
spectral-function expressions derived here are used in Ref. \cite{V} in the evaluation of
closed-form expressions for the one-electron and two-electron spectral-weight
distributions of the model metallic phase. The 1D Hubbard model is one of the few
realistic models for which one can exactly calculate all the energy eigenstates and their
energies \cite{Lieb,Takahashi}. In addition to the applications to the study of the
unusual properties of the quasi-1D compounds presented in Refs. \cite{super,spectral1},
our results are also of interest for the understanding of the spectral properties of the
new quantum systems described by ultra-cold fermionic atoms in optical lattices with
on-site repulsion \cite{Zoller}.

The electron - rotated-electron unitary transformation \cite{Harris} used in Ref.
\cite{I} for all values of the on-site repulsion $U$ and the pseudofermion scattering
theory introduced in Ref. \cite{S} play a central role in the construction of the {\it
pseudofermion dynamical theory} introduced here. For finite values of $U$ very little is
known about the finite-energy spectral properties of the model. This is in contrast to
simpler models \cite{Lee}. Unfortunately, combination of the model Bethe-ansatz solution
\cite{Lieb,Takahashi} with bosonization, conformal-field theory, or g-ology and
Renormalization Group \cite{Schulz,Karlo} only allows the derivation of low-energy
correlation-function expressions. In the limit of infinite $U$ the spectral functions can
be evaluated by the method presented in Ref. \cite{Penc97} and there are recent numerical
results for finite values of $U$ \cite{Eric}, but it is difficult to extract from them
information about the microscopic processes that control the unusual spectral properties
of the model.

The paper is organized as follows: In Sec. \ref{SecII} we introduce the model and the
spectral-function problem and summarize the pseudofermion operational description used in
our study. In Sec. \ref{SecIII} we write the general spectral functions in terms of
rotated-electron operators. The description of the rotated-electron elementary processes
in terms of pseudofermion operators is the problem addressed in Sec. \ref{SecIV}. In Sec.
\ref{SecV} we express the spectral functions as a convolution of pseudofermion spectral
functions and study the pseudofermion determinants involved in the expressions of these
functions. Finally, the concluding remarks are presented in Sec. \ref{SecVI}.

\section{THE MODEL, THE SPECTRAL FUNCTIONS, AND THE PSEUDOFERMION DESCRIPTION}
\label{SecII}

In a chemical potential $\mu $ and magnetic field $H$ the 1D Hubbard Hamiltonian can be
written as,
\begin{equation}
\hat{H} = {\hat{H}}_{SO(4)} + \sum_{\alpha=c,\,s}\mu_{\alpha }\,{\hat{S}}^{\alpha}_z \, ;
\hspace{1cm} {\hat{H}}_{SO(4)} = -t\sum_{j =1}^{N_a}\sum_{\sigma =\uparrow
,\downarrow}\sum_{\delta =-1,+1} c_{j,\,\sigma}^{\dag}\,c_{j+\delta,\,\sigma} +
U\,\sum_{j =1}^{N_a} [\hat{n}_{j,\,\uparrow}-1/2][\,\hat{n}_{j,\,\downarrow}-1/2] \, .
\label{H}
\end{equation}
Here the operators $c_{j,\,\sigma }^{\dagger }$ and $c_{j,\,\sigma}$ are the
spin-projection $\sigma $ electronic creation and annihilation operators at site $j$ and
$\hat{n}_{j,\,\sigma}= c_{j,\,\sigma }^{\dagger }\,c_{j,\,\sigma}$ where $j=1,2,...,N_a$.
The number of lattice sites $N_a$ is even and very large. We consider periodic boundary
conditions. In the first expression of Eq. (\ref{H}), $\mu_c=2\mu$, $\mu_s=2\mu_0 H$,
$\mu_0$ is the Bohr magneton, and the diagonal generators of the $\eta$-spin and spin
$SU(2)$ algebras \cite{HL,Yang89} ${\hat{S }}^c_z$ and ${\hat{S }}^s_z$, respectively,
are given in Eq. (2) of Ref. \cite{I}. The Hamiltonian ${\hat{H}}_{SO(4)}$ of Eq.
(\ref{H}) commutes with the six generators of these two algebras, their off-diagonal
generators being given in Eqs. (7) and (8), respectively, of Ref. \cite{I}. The electron
number operator reads ${\hat{N}}=\sum_{\sigma} \hat{N}_{\sigma}$, where
${\hat{N}}_{\sigma}= \sum_{j=1}^{N_a}\hat{n}_{j,\,\sigma}$. For simplicity, we use units
such that the Planck constant and electronic lattice constant are one. The model
(\ref{H}) describes $N_{\uparrow}$ spin-up electrons and $N_{\downarrow}$ spin-down
electrons in a chain of $N_a$ sites, whose length in the units used here reads $L=N_a$.
We introduce the Fermi momenta which, in the thermodynamic limit $L\rightarrow\infty$,
are given by $\pm k_{F\sigma}=\pm \pi n_{\sigma }$ and $\pm k_F=\pm [k_{F\uparrow}+
k_{F\downarrow}]/2=\pm \pi n/2$, where $n_{\sigma}=N_{\sigma}/L$ and $n=N/L$. The
electronic density can be written as $n=n_{\uparrow }+n_{\downarrow}$ and the spin
density is given by $m=n_{\uparrow}-n_{\downarrow}$. We denote the $\eta$-spin value
$\eta$ and projection $\eta_z$ (and spin value $S$ and projection $S_z$) of an energy
eigenstate by $S_c$ and $S_c^z$ (and $S_s$ and $S_s^z$), respectively. The momentum
operator reads,
\begin{equation}
\hat{P} = \sum_{\sigma=\uparrow ,\,\downarrow }\sum_{k}\, \hat{N}_{\sigma} (k)\, k =
{L\over 2\pi}\sum_{\sigma=\uparrow ,\,\downarrow }\,\int_{-\pi}^{+\pi} dk\,
\hat{N}_{\sigma} (k)\, k \, , \label{Popel}
\end{equation}
and commutes with the Hamiltonians given in Eq. (\ref{H}). The spin-projection $\sigma$
momentum distribution operator appearing in Eq. (\ref{Popel}) is given by
$\hat{N}_{\sigma} (k) = c_{k,\,\sigma }^{\dagger }\,c_{k,\,\sigma }$. Here $c_{k,\,\sigma
}^{\dagger }$ and $c_{k,\,\sigma}$ are the spin-projection $\sigma $ electron creation
and annihilation operators of momentum $k$. These operators are related to the above
local operators as follows,
\begin{equation}
c_{k,\,\sigma}^{\dag} = {1\over{\sqrt{N_a}}}\sum_{j=1}^{N_a}\,e^{+ikja}\,
c_{j,\,\sigma}^{\dag} \, ; \hspace{1cm} c_{k,\,\sigma} =
{1\over{\sqrt{N_a}}}\sum_{j=1}^{N_a}\,e^{-ikja}\, c_{j,\,\sigma} \, . \label{ck}
\end{equation}

The Bethe-ansatz solvability of the 1D Hubbard model is restricted to the Hilbert
subspace spanned by {\it regular states}, {\it i.e.} the lowest-weight states (LWSs) of
the $\eta$-spin and spin algebras such that $S_{\alpha}= -S^{\alpha}_z$, where $\alpha
=c,\,s$ \cite{I}. For simplicity, in this paper we restrict our considerations to values
of the electronic density $n$ and spin density $m$ such that $0\leq n \leq 1$ and $0\leq
m \leq n$, respectively. Often our expressions are different for the $n=1$ Mott-Hubbard
insulator phase and $0< n < 1$ metallic phase (and for $m=0$ zero spin density and $0< m
< n$ finite spin densities).

The main aim of this paper is the evaluation of expressions for finite-$\omega$
$\cal{N}$-electron spectral functions $B_{{\cal{N}}}^{l} (k,\,\omega)$, such that $l=\pm
1$, of the general form,
\begin{equation}
B_{{\cal{N}}}^{l} (k,\,\omega) = \sum_{f}\, \vert\langle f\vert\,
{\hat{O}}_{{\cal{N}}}^{l} (k) \vert GS\rangle\vert^2\,\delta\Bigl(\omega - l[E_f -
E_{GS}]\Bigr) \, ; \hspace{0.5cm} l\omega > 0 \, ; \hspace{0.5cm} l = \pm 1 \, ,
\label{ABON}
\end{equation}
where the operators in the matrix elements are such that,
\begin{equation}
{\hat{O}}_{{\cal{N}}}^{+1} (k) \equiv {\hat{O}}_{{\cal{N}}}^{\dag} (k) \, ; \hspace{1cm}
{\hat{O}}_{{\cal{N}}}^{-1} (k) \equiv {\hat{O}}_{{\cal{N}}} (k) \, . \label{O-notation}
\end{equation}
Here the $f$ summation runs over the excited energy eigenstates, the energies $E_f$
correspond to these states, $E_{GS}$ is the ground-state energy, and we use a momentum
extended scheme such that $k\in (-\infty,\,+\infty)$. The operators
${\hat{O}}_{{\cal{N}}}^{\dag} (k)$ and ${\hat{O}}_{{\cal{N}}} (k)$ carry momentum $k$ and
are denoted in Eq. (\ref{O-notation}) by ${\hat{O}}_{{\cal{N}}}^l (k)$ where $l=+1$ and
$l=-1$, respectively. They are related to the local operators
${\hat{O}}_{{\cal{N}},\,j}^{\dag}\equiv {\hat{O}}_{{\cal{N}},\,j}^{+1}$ and
${\hat{O}}_{{\cal{N}},\,j}\equiv {\hat{O}}_{{\cal{N}},\,j}^{-1}$, respectively, by a
Fourier transform.

The local operators ${\hat{O}}_{{\cal{N}},\,j}^{l}$ can be written as a product of
\begin{equation}
{\cal{N}}=\sum_{l_c,\,l_s=\pm 1} {\cal{N}}_{l_c,\,l_s}^l \, ; \hspace{0.5cm} l = \pm 1 \,
, \label{d}
\end{equation}
local electronic creation and annihilation operators. Here ${\cal{N}}_{l_c,\,l_s}^l$ is
the number of local electronic creation and annihilation operators of the operator
${\hat{O}}_{{\cal{N}},\,j}^{l}$ for $l_c=-1$ and $l_c=+1$, respectively, and with spin
down and spin up for $l_s=-1$ and $l_s=+1$, respectively. It is assumed that the ratio
${\cal{N}}/N_a$ vanishes in the thermodynamic limit. Note that, by construction,
${\cal{N}}_{l_c,\,l_s}^{+1}$ and ${\cal{N}}_{l_c,\,l_s}^{-1}$ are such that
${\cal{N}}_{l_c,\,l_s}^{+1}= {\cal{N}}_{-l_c,\,l_s}^{-1}$, ${\cal{N}}_{-1,\,l_s}^{+1}\geq
{\cal{N}}_{+1,\,l_s}^{+1}$, and ${\cal{N}}_{-1,\,l_s}^{-1}\leq
{\cal{N}}_{+1,\,l_s}^{-1}$. For ${\cal{N}}>1$ the operator
${\hat{O}}_{{\cal{N}},\,j}^{l}$ has a well defined local structure involving the
${\cal{N}}_{-1,\,l_s}$ electronic creation operators of spin projection $l_s/2$, and
${\cal{N}}_{+1,\,l_s}^l$ electronic annihilation operators of spin projection $l_s/2$
located in neighboring lattice sites. The more usual cases for the description of
experimental studies correspond to ${\cal{N}}=1$ and ${\cal{N}}=2$. Examples of
${\cal{N}}$-electron operators ${\hat{O}}_{{\cal{N}}} (k) \equiv
{\hat{O}}_{{\cal{N}}}^{-1}(k)$ are the one-electron operator ${\hat{O}}_{1} (k) =
c_{k,\,\sigma}$ (measured in the angle-resolved photoelectron spectroscopy), the
spin-projection $\sigma$ density operator ${\hat{O}}_{2}^{\sigma sd} (k) = {1\over
\sqrt{N_a}}\sum_{k'}c^{\dagger}_{k+k',\,\sigma} c_{k',\,\sigma}$, the transverse
spin-density operator ${\hat{O}}_{2}^{sdw} (k) = {1\over
\sqrt{N_a}}\sum_{k'}c^{\dagger}_{k+k',\,\uparrow} c_{k',\,\downarrow}$, the on-site
s-wave singlet superconductivity operator ${\hat{O}}_2^{oss} (k) = {1\over
\sqrt{N_a}}\sum_{k'} c_{k-k',\,\uparrow}c_{k',\,\downarrow}$, and the spin-projection
$\sigma$ triplet superconductivity operator ${\cal{O}}_2^{\sigma ts} (k) = {1\over
\sqrt{N_a}}\sum_{k'}\cos (k')\,c_{k-k',\,\sigma}c_{k',\,\sigma}$. The corresponding local
operators ${\hat{O}}_{{\cal{N}},\,j}^{-1}\equiv {\hat{O}}_{{\cal{N}},\,j}$ are
${\hat{O}}_{1,\,j} = c_{j,\,\sigma}$, ${\hat{O}}_{2,\,j}^{\sigma sd} =
c^{\dagger}_{j,\,\sigma} c_{j,\,\sigma}$, ${\hat{O}}_{2,\,j}^{sdw} =
c^{\dagger}_{j,\,\uparrow} c_{j,\,\downarrow}$, ${\hat{O}}_{2,\,j}^{oss} =
c_{j,\,\uparrow}c_{j,\,\downarrow}$, and ${\hat{O}}_{2,\,j}^{\sigma ts}=
c_{j,\,\sigma}c_{j+1,\,\sigma}$, respectively. The charge density operator (measured in
density-density electron energy loss spectroscopy and inelastic X-ray scattering) is
written in terms of the above spin-up and spin-down density operators. The operators
${\hat{O}}_{{\cal{N}}}^{l} (k)$ of physical interest, correspond in general to operators
${\hat{O}}_{{\cal{N}},\,j}^{l}$ whose ${\cal{N}}$ elementary electronic operators create
or annihilate electrons in a compact domain of lattice sites. For instance, if
${\hat{O}}_{2,\,j}^{+1}=c^{\dag }_{j,\,\downarrow}\,c^{\dag }_{j+i,\,\uparrow}$ and thus
${\hat{O}}_{2,\,j}^{-1}=c_{j+i,\,\uparrow}\,c_{j,\,\downarrow}$, the interesting cases
correspond to $i=0$ (on-site s-wave singlet superconductivity) and $i=1$ (extended s-wave
singlet superconductivity).

The $k$ dependence of the spectral functions (\ref{ABON}) can be transferred from the
${\cal{N}}$-electron operators ${\hat{O}}_{{\cal{N}}}^{l} (k)$ to the excited energy
eigenstates as follows,
\begin{equation}
B_{{\cal{N}}}^{l} (k,\,\omega) = \sum_{f}\, N_a\vert\langle f\vert\,
{\hat{O}}_{{\cal{N}},\,0}^{l}\vert GS\rangle\vert^2\,\delta\Bigl( \omega - l[E_f -
E_{GS}]\Bigr)\,\delta_{k,\,l[k_f-k_{GS}]} \, ; \hspace{0.5cm} l\omega > 0 \, ;
\hspace{0.5cm} l=\pm 1 \, . \label{ABONj}
\end{equation}
Here, ${\hat{O}}_{{\cal{N}},\,0}^{l}$ is the $j=0$ local operator
${\hat{O}}_{{\cal{N}},\,j}^{l}$ considered above, $k_f$ is the momentum of the excited
energy eigenstates, and $k_{GS}$ denotes the ground-state momentum. In this expression,
we have chosen $j=0$ for the local operator ${\hat{O}}_{{\cal{N}},\,j}^{l}$. Due to
translational invariance, the value of the functions (\ref{ABONj}) is independent of this
special choice.

Let us summarize the basic information about the holon, spinon, pseudoparticle, and
pseudofermion descriptions needed for our studies. (For further information, see Refs.
\cite{V,I,IV,II}.) These studies involve the electron - rotated-electron unitary
transformation, such that rotated-electron double occupancy is a good quantum number for
all $U/t$ values \cite{I}. As the Fermi-liquid quasiparticles, the rotated electrons have
the same charge and spin as the electrons, but refer to all energies and reorganize in
terms of $[N_a-N_c]$ $\eta$-spin $1/2$ holons, $N_c$ spin $1/2$ spinons, and $N_c$
spinless and $\eta$-spinless $c$ pseudoparticles, where $N_c$ is the number of
rotated-electron singly occupied sites \cite{I}. We use the notation $\pm 1/2$ holons and
$\pm 1/2$ spinons, which refers to the $\eta$-spin and spin projections, respectively.
The $\pm 1/2$ holons of charge $\pm 2e$ correspond to rotated-electron unoccupied $(+)$
and doubly-occupied $(-)$ sites. The complex behavior occurs for the spin-projection
$\sigma$-rotated electrons occupying singly occupied sites: their spin degrees of freedom
originate chargeless spin-projection $\sigma$ spinons, whereas their charge part gives
rise to $\eta$-spinless and spinless $c$ pseudoparticles of charge $-e$.

Based on symmetry considerations, we can classify the $\pm 1/2$ holons and $\pm 1/2$
spinons into two classes: those which remain invariant under the electron -
rotated-electron unitary transformation, and those which do not. The former are called
independent $\pm 1/2$ holons and independent $\pm 1/2$ spinons. For instance, the  $\pm
1/2$ Yang holons and $\pm 1/2$ HL spinons \cite{V,I,IV,II} with numbers reading
$L_{c,\,\pm 1/2}=[S_c\mp S_c^z]$ and $L_{s,\,\pm 1/2}=[S_s\mp S_s^z]$, respectively,
belong to the former group of holons and spinons. The latter are part of $\eta$-spin-zero
$2\nu$-holon composite $c\nu$ pseudoparticles and spin-zero $2\nu$-spinon composite
$s\nu$ pseudoparticles, respectively, where $\nu=1,2,...$ is the number of $+1/2$ and
$-1/2$ holon or $+1/2$ and $-1/2$ spinon pairs. Thus, the total number of $\pm 1/2$
holons $(\alpha =c)$ and $\pm 1/2$ spinons $(\alpha =s)$ reads $M_{\alpha,\,\pm
1/2}=L_{\alpha,\,\pm 1/2}+\sum_{\nu =1}^{\infty}\nu\,N_{\alpha\nu}$, where
$N_{\alpha\nu}$ denotes the number of composite $\alpha\nu$ pseudoparticles. The total
number of holons $(\alpha =c)$ and spinons $(\alpha =s)$ is then given by
$M_{\alpha}=L_{\alpha}+2\sum_{\nu =1}^{\infty}\nu\,N_{\alpha\nu}$ where
$L_{\alpha}=2S_{\alpha}$ denotes the total number of Yang holons $(\alpha =c)$ and HL
spinons $(\alpha =s)$. These numbers are such that $M_c=[N_a-N_c]$ and $M_s=N_c$. The
pseudoparticles can be defined in terms of bare-momentum or spatial coordinates
\cite{IV}. In addition to the Yang holons and HL spinons, also the holons and spinons
associated with $\alpha\nu\neq c0,\, s1$ pseudoparticles of limiting bare-momentum values
$\pm q_{\alpha\nu}$ are independent holons and spinons. ($q_{\alpha\nu}$ is given in Eq.
(B.14) of Ref. \cite{I}.) Indeed, the invariance under the electron - rotated-electron
unitary transformation of such $c\nu$ pseudoparticles (and $s\nu$ pseudoparticles)
implies that they separate into $2\nu$ independent holons (and $2\nu$ independent
spinons) and a $c\nu$ (and $s\nu$) FP scattering center \cite{V}. (These centers are
defined in Ref. \cite{V}.) The emergence of the exotic quantum phases of matter
considered in our study involves a second unitary transformation, which maps the $c$
pseudoparticles (and composite $\alpha\nu$ pseudoparticles) onto $c$ pseudofermions (and
composite $\alpha\nu$ pseudofermions) \cite{V}. Such a transformation introduces shifts
of order $1/L$ in the pseudoparticle discrete momentum values and leaves all other
pseudoparticle properties invariant. Here we use the designation $c0$ pseudoparticle and
pseudofermion for the $c$ pseudoparticle and pseudofermion, respectively. Thus, the
$c\nu$ and $s\nu$ branches are such that $\nu =0,1,2,...$ and $\nu =1,2,...$,
respectively.

The local $\alpha\nu$ pseudofermion creation (and annihilation) operator $f^{\dag
}_{x_j,\,\alpha\nu}$ (and $f_{x_j,\,\alpha\nu}$) creates (and annihilates) a $\alpha\nu$
pseudofermion at the effective $\alpha\nu$ lattice site of spatial coordinate $x_j
=j\,a^0_{\alpha\nu}$. Here $j=1,2,...,N^*_{\alpha\nu}$ and
$a^0_{\alpha\nu}=L/N^*_{\alpha\nu}=N_a/N^*_{\alpha\nu}$ is the effective $\alpha\nu$
lattice constant given in Eq. (55) of Ref. \cite{IV} in units of the electronic lattice
constant. The general expression of the number of effective $\alpha\nu$ lattice sites
$N^*_{\alpha\nu}$ is given in Eq. (B6) of Ref. \cite{I}, where the number of $\alpha\nu$
pseudofermion holes $N^h_{\alpha\nu}$ is provided in Eq. (B.11) of the same reference.
(The number of pseudofermion and pseudofermion holes equals that of the corresponding
pseudoparticle and pseudoparticle holes \cite{V,IV}.) The operator $f^{\dag
}_{x_j,\,\alpha\nu}$ (and $f_{x_j,\,\alpha\nu}$) is related to the operator
$f^{\dag}_{{\bar{q}}_j,\,\alpha\nu}$ (and $f_{{\bar{q}}_j,\,\alpha\nu}$), which refers to
$\alpha\nu$ pseudofermions of canonical-momentum ${\bar{q}}_j$, by a Fourier transform.
The discrete canonical-momentum values of the $\alpha\nu$ pseudofermions have a
functional character and read \cite{V,S},
\begin{equation}
{\bar{q}}_j = q_j + Q^{\Phi}_{\alpha\nu} (q_j)/L = [2\pi/ L] I^{\alpha\nu}_j +
Q^{\Phi}_{\alpha\nu} (q_j)/L \, ; \hspace{0.5cm} j = 1, 2, ..., N^*_{\alpha\nu} \, ,
\label{bar-q}
\end{equation}
where $q_j=[2\pi I^{\alpha\nu}_j]/L$ \cite{I} is the bare-momentum carried by the
$\alpha\nu$ pseudoparticles. Here $I^{\alpha\nu}_j$ are the actual quantum numbers
provided by the Bethe-ansatz solution \cite{I}. Although the $\alpha\nu$ pseudoparticles
carry bare-momentum $q_j$, one can also label the corresponding $\alpha\nu$
pseudofermions by such a bare-momentum. When one refers to the pseudofermion
bare-momentum $q_j$, one means that $q_j$ is the bare-momentum value that corresponds to
the canonical momentum ${\bar{q}}_j = q_j + Q^{\Phi}_{\alpha\nu} (q_j)/L$. Here and in
Eq. (\ref{bar-q}) $Q^{\Phi}_{\alpha\nu}(q_j)/2$ is a $\alpha\nu$ pseudofermion overall
scattering phase shift given by \cite{S},
\begin{equation}
Q^{\Phi}_{\alpha\nu} (q_j)/2 = \pi \sum_{\alpha'\nu'}\,
\sum_{j'=1}^{N^*_{\alpha'\nu'}}\,\Phi_{\alpha\nu,\,\alpha'\nu'}(q_j,q_{j'})\, \Delta
N_{\alpha'\nu'}(q_{j'}) \, ; \hspace{0.5cm} j = 1, 2, ..., N^*_{\alpha\nu} \, ,
\label{qcan1j}
\end{equation}
where $\Delta N_{\alpha\nu}(q_{j})=\Delta {\cal{N}}_{\alpha\nu} ({\bar{q}}_j)$ is the
distribution function deviation $\Delta N_{\alpha\nu} (q_j) = N_{\alpha\nu} (q_j) -
N^0_{\alpha\nu} (q_j)$. The canonical-momentum distribution function
${\cal{N}}_{\alpha\nu}({\bar{q}}_j)$ (and bare-momentum distribution function
$N_{\alpha\nu} (q_j)$) is given by ${\cal{N}}_{\alpha\nu}({\bar{q}}_j)=1$ and
${\cal{N}}_{\alpha\nu}({\bar{q}}_j)=0$ (and $N_{\alpha\nu} (q_j)=1$ and $N_{\alpha\nu}
(q_j)=0$) for pseudofermions and pseudofermion holes (and pseudoparticles and
pseudoparticle holes), respectively \cite{V}. The ground-state densely-packed
bare-momentum distribution function $N^0_{\alpha\nu} (q_j)$ is defined in Eqs.
(C.1)-(C.3) of Ref. \cite{I}. The $\alpha\nu\neq c0,\,s1$ pseudofermion limiting
canonical-momentum values play an important role in the theory and read,
\begin{equation}
q_{\alpha\nu}^0 = q_{\alpha\nu} + Q^{\Phi}_{\alpha\nu} (q_{\alpha\nu})/L  \, ;
\hspace{0.5cm} \alpha\nu\neq c0,\,s1 \, , \label{limiting}
\end{equation}
where $q_{\alpha\nu}^0$ is the ground-state limiting bare-momentum value given in Eqs.
(C.13) and (C.14) of Ref. \cite{I} and $q_{\alpha\nu}$ the excited-energy-eigenstate
limiting bare-momentum value provided in Eq. (B.14) of the same reference. In contrast to
the $\alpha\nu$ pseudoparticles, the $\alpha\nu$ pseudofermions have no
residual-interaction energy terms \cite{V}. Instead, under the ground-state -
excited-energy-eigenstate transitions the $\alpha\nu$ pseudofermions and $\alpha\nu$
pseudofermion holes undergo elementary scattering events with the $\alpha'\nu'$
pseudofermions and $\alpha'\nu'$ pseudofermion holes created in these transitions
\cite{S}. This leads to the elementary two-pseudofermion phase shifts
$\pi\,\Phi_{\alpha\nu,\,\alpha'\nu'}(q_j,{q'}_j)$ on the right-hand side of the overall
scattering phase shift (\ref{qcan1j}), which are defined by a set of integral equations
\cite{V,S}. The overall $\alpha\nu$ pseudofermion or hole phase shift,
\begin{equation}
Q_{\alpha\nu}(q_j)/2 = Q_{\alpha\nu}^0/2 + Q^{\Phi}_{\alpha\nu} (q_j)/2 \, ,
\label{Qcan1j}
\end{equation}
plays an important role in the pseudofermion theory \cite{V,S}. Here $Q_{\alpha\nu}^0/2$
can have the values $Q_{\alpha\nu}^0/2 =0,\,\pm\pi/2$ \cite{V,S}.

The pseudofermion description refers to a Hilbert subspace called {\it pseudofermion
subspace} (PS) where the $\cal{N}$-electron excitations are contained \cite{V,S}. The PS
is spanned by the initial ground state and the excited energy eigenstates generated from
it by the following types of processes (A)-(C), which are defined in more detail in Ref.
\cite{V}: (A) finite-energy and finite-momentum elementary $c0$ and $s1$ pseudofermion
processes plus creation of $\alpha\nu\neq c0,\,s1$ pseudofermions with bare-momentum
values $q\neq \pm q_{\alpha\nu}$, (B) zero-energy and finite-momentum processes that
change the number of $c0$ and $s1$ pseudofermions at their {\it Fermi points}, which for
the ground state and $L\rightarrow\infty$ read,
\begin{equation}
q^0_{Fc0} = 2k_F \, ; \hspace{0.5cm} q^0_{Fs1} = k_{F\downarrow} \, , \label{q0Fcs}
\end{equation}
plus creation of independent $-1/2$ holons and/or $-1/2$ spinons, and (C) low-energy and
small-momentum elementary $c0$ and $s1$ pseudofermion particle-hole processes in the
vicinity of the {\it Fermi points}. The PS contains subspaces of several CPHS ensemble
subspaces. (Here CPHS stands for $c0$ pseudofermion, holon, and spinon.) A CPHS ensemble
subspace is spanned by all energy eigenstates with fixed values for the $-1/2$ Yang holon
number $L_{c,\,-1/2}$, $-1/2$ HL spinon number $L_{s,\,-1/2}$, $c0$ pseudofermion number
$N_{c0}$, and for the sets of composite $\alpha\nu$ pseudofermion numbers $\{N_{c\nu}\}$
and $\{N_{s\nu}\}$ corresponding to the $\nu=1,2,...$ branches.

The pseudofermion bare-momentum dependent energy dispersions $\epsilon_{c0} (q)$,
$\epsilon_{s1} (q)$, $\epsilon_{c\nu} (q) = 2\nu\mu+\epsilon_{c\nu}^0 (q)$ for $\nu>0$,
and $\epsilon_{s\nu} (q) = 2\nu\mu_0 H+\epsilon_{s\nu}^0 (q)$ for $\nu>1$, where $\mu
=\mu (n)$ and $H=H(m)$ correspond to the density and magnetization curves, respectively,
are defined and studied in Refs. \cite{I,V,II}. Such energy dispersions play a crucial
role in the expressions of the $\cal{N}$-electron spectral functions. For $m=0$, the
energy $2\mu$ is an increasing function of $U$ and a decreasing function of the density
$n$ with the following limiting values,
\begin{equation}
2\mu = 4t\cos (\pi na/2) \, , \hspace{0.3cm} U/t\rightarrow 0 \, ; \hspace{0.5cm} U+
4t\cos (\pi na)  \, , \hspace{0.3cm} U/t\rightarrow\infty \, ; \hspace{0.5cm} U + 4t \, ,
\hspace{0.3cm} n\rightarrow 0 \, ; \hspace{0.5cm} E_{MH}  \, , \hspace{0.3cm}
n\rightarrow 1 \, , \label{Eu}
\end{equation}
where $E_{MH}$ is the half-filling Mott-Hubbard gap \cite{Lieb}.

The evaluation of matrix elements between energy eigenstates considered in Sec. V
involves pseudofermion operators $f^{\dag }_{{\bar{q}},\,\alpha\nu}$ and
$f_{{\bar{q}}',\,\alpha\nu}$ such that the canonical momentum values ${\bar{q}}$ and
${\bar{q}}'=q'$ correspond to an excited-energy-eigenstate and initial ground-state CPHS
ensemble subspaces, respectively. In that case the unusual pseudofermion anticommutation
relations read \cite{V,S},
\begin{equation}
\{f^{\dag }_{{\bar{q}},\,\alpha\nu},\,f_{q',\,\alpha' ,\,\nu'}\} =
\delta_{\alpha\nu,\,\alpha'\nu'}\,{1\over N^*_{\alpha\nu}}\,e^{-i({\bar{q}}-q')/
2}\,e^{iQ_{\alpha\nu}(q)/2}\,{\sin\Bigl(Q_{\alpha\nu} (q)/ 2\Bigr)\over\sin
([{\bar{q}}-q']/2)} \, , \label{pfacrGS}
\end{equation}
and $\{f^{\dag }_{{\bar{q}},\,\alpha\nu},\,f^{\dag }_{q',\,\alpha'
,\,\nu'}\}=\{f_{{\bar{q}},\,\alpha\nu},\,f_{q',\,\alpha' ,\,\nu'}\} =0$.

\section{SPECTRAL FUNCTIONS IN TERMS OF ROTATED-ELECTRON OPERATORS}
\label{SecIII}

Here we express the general $\cal{N}$-electron spectral functions (\ref{ABONj}) in terms
of rotated-electron creation and annihilation operators and evaluate the spectral-weight
contributions from the Yang holons and HL spinons. Our first goal is identifying the set
of CPHS ensemble subspaces which are spanned by the excited energy eigenstates generated
by application onto the initial ground state of the operator
${\hat{O}}_{{\cal{N}},\,0}^{l}$ of Eq. (\ref{ABONj}). For clarification of this problem,
we must find the set of deviation numbers $\Delta N_{c0}$, $\Delta N_{s1}$, $\{\Delta
L_{\alpha,\,-1/2}\}$, and $\{\Delta N_{\alpha\nu}\}$ for $\alpha\nu \neq c0,\,s1$ that
are generated by application onto the ground state of that operator. According to the
results of Refs. \cite{V,I}, for the ground state
$M_{c,\,-1/2}=L_{\alpha,\,-1/2}=N_{\alpha\nu}=0$ for the $\alpha\nu \neq c0,\,s1$
branches and thus $\Delta M_{c,\,-1/2} =M_{c,\,-1/2}$, $\Delta
L_{\alpha,\,-1/2}=L_{\alpha,\,-1/2}$, and $\Delta N_{\alpha\nu} =N_{\alpha\nu}$ for the
latter branches.

First, we note that the values of the $+1/2$ holon and $+1/2$ spinon number deviations
are such that,
\begin{equation}
\Delta M_{c,\,+1/2} = - \Delta N_{c0} - M_{c,\,-1/2} \, ; \hspace{0.5cm} \Delta
M_{s,\,+1/2} = \Delta N_{c0} - \Delta M_{s,\,-1/2} \label{M+NcM-} \, ,
\end{equation}
and thus are dependent on the values of the $-1/2$ holon and $-1/2$ spinon numbers and
$c0$ pseudofermion number deviations. Also the occupancy configurations of the $-1/2$
holons and $-1/2$ spinons determine those of the $+1/2$ holons and $+1/2$ spinons.
Indeed, the $-1/2$ holons and $+1/2$ holons correspond to the rotated-electron
doubly-occupied sites and unoccupied sites, respectively, of a charge sequence \cite{IV}.
The point is that the spatial position of the unoccupied sites corresponds to the sites
left over by the rotated-electron doubly occupied sites of a charge sequence. The same
applies to the $-1/2$ spinons and $+1/2$ spinons, provided that we replace the
rotated-electron doubly-occupied sites and unoccupied sites by sites singly occupied by
spin-down and spin-up rotated electrons, respectively, and the charge sequence by the
spin sequence \cite{IV}. Moreover, the values of the corresponding $+1/2$ Yang holon and
$+1/2$ HL spinon number deviations read,
\begin{equation}
\Delta L_{c,\,+1/2} = - \Delta N_{c0} - 2\sum_{\nu =1}^{\infty} \nu\,N_{c,\,\nu} -
L_{c,\,-1/2} \, , \label{ScNN}
\end{equation}
and
\begin{equation}
\Delta L_{s,\,+1/2} = \Delta N_{c0} - 2 \Delta N_{s1} - 2\sum_{\nu =2}^{\infty}
\nu\,N_{s,\,\nu} - L_{s,\,-1/2} \, , \label{ScsNN}
\end{equation}
respectively, and thus are not independent. One does not need to provide these values in
order to specify a CPHS ensemble subspace. Therefore, often we do not consider in the
expressions below the values of the holon numbers $M_{c,\,+1/2}$ and $L_{c,\,+1/2}$ and
of the spinon numbers $M_{s,\,+1/2}$ and $L_{s,\,+1/2}$.

The values of the deviations $\Delta N_{\uparrow}$ and $\Delta N_{\downarrow}$ specific
to a given $\cal{N}$-electron operator, lead to sum rules for the values of the number
deviations of pseudofermions, $-1/2$ Yang holons, and $-1/2$ HL spinons as follows,
\begin{equation}
\Delta N = \Delta N_{c0} + 2L_{c,\,-1/2} + 2\sum_{\nu =1}^{\infty}\nu\,N_{c\nu} \, ,
\label{DNu}
\end{equation}
and
\begin{equation}
\Delta (N_{\downarrow} - N_{\uparrow})= 2\Delta N_{s1} -\Delta N_{c0} + 2L_{s,\,-1/2} +
2\sum_{\nu =2}^{\infty}\nu\,N_{s\nu} \, . \label{DNd}
\end{equation}
Only transitions to excited energy eigenstates associated with deviations obeying the sum
rules (\ref{DNu}) and (\ref{DNd}) are permitted. The same deviations are associated with
sum rules obeyed by the numbers ${\cal{N}}_{l_c,\,l_s}^{l}$ of Eq. (\ref{d}) for the
operator ${\hat{O}}_{{\cal{N}},\,j}^{l}$ appearing in the general spectral-function
expressions of Eq. (\ref{ABONj}). Such sum rules read,
\begin{equation}
\Delta N = \sum_{l_c,\,l_s=\pm 1}(-l_c)\,{\cal{N}}_{l_c,\,l_s}^l \, ; \hspace{1cm} \Delta
(N_{\downarrow}-N_{\uparrow}) = \sum_{l_c,\,l_s=\pm
1}(l_c\,l_s)\,{\cal{N}}_{l_c,\,l_s}^{l} \, . \label{NupdodllO}
\end{equation}

Furthermore, it is straightforward to show that the following selection rule is valid for
initial ground states corresponding to the density values considered in this paper: The
values of the numbers of $-1/2$ Yang holons and $-1/2$ HL spinons generated by
application onto the ground state of the $\cal{N}$-electron operator
${\hat{O}}_{{\cal{N}},\,j}^{l}$, Eq. (\ref{ABONj}), are restricted to the following
ranges,
\begin{equation}
L_{c,\,-1/2} =0,1,2,...,\sum_{l_s=\pm 1}{\cal{N}}_{-1,\,l_s}^{l} \, ; \hspace{1cm}
L_{s,\,-1/2} =0,1,2,..., \sum_{l_c,\,l_s=\pm 1}\delta_{l_c,\,l_s}\,
{\cal{N}}_{l_c,\,l_s}^{l} \, ; \hspace{0.5cm} l = \pm 1 \, , \label{srcsL}
\end{equation}
respectively. Here the numbers ${\cal{N}}_{l_c,\,l_s}^{l}$ are those of Eq. (\ref{d})
specific to that operator.

Further selection rules in terms of the rotated-electron expressions for the operator
${\hat{O}}_{{\cal{N}},\,0}^{l}$ of the general spectral-function (\ref{ABONj}) are given
in the ensuing section.

Let us label the excited energy eigenstates of the state summations of the general
$\cal{N}$-electron spectral function (\ref{ABONj}) according to their CPHS ensemble
subspace. (We recall that all excited energy eigenstates belonging to a given CPHS
ensemble subspace have the same values for the set of deviation numbers $\Delta N_{c0}$
and $\Delta N_{s1}$ and numbers $\{L_{\alpha,\,-1/2}\}$  for $\alpha= c,\,s$, and
$\{N_{\alpha\nu}\}$ for the $\alpha\nu\neq c0,\,s1$ branches.) This procedure leads to
the following expression for the spectral function (\ref{ABONj}),
\begin{equation}
B_{{\cal{N}}}^{l} (k,\,\omega) = \sum_{\{\Delta
N_{\alpha\nu}\},\,\{L_{\alpha,\,-1/2}\}}\sum_{f}\, N_a\vert\langle f;\,C\vert\,
{\hat{O}}_{{\cal{N}},\,0}^{l} \vert GS\rangle\vert^2\,\delta\Bigl(\omega - l\Delta
E_{cphs}\Bigr)\,\delta_{k,\,l\Delta k_{cphs}} \, ; \hspace{0.5cm} l\omega > 0 \, ,
\label{ABONj-HS}
\end{equation}
where $l=\pm 1$ and the summation $\sum_{\{\Delta
N_{\alpha\nu}\},\,\{L_{\alpha,\,-1/2}\}}$ runs over the CPHS ensemble subspaces whose
deviation values obey the sum rules (\ref{DNu}) and (\ref{DNd}) and selection rules
(\ref{srcsL}). Moreover, the summation $\sum_{f}$ runs over the excited energy
eigenstates $\vert f;\,C\rangle$ of a given CPHS ensemble subspace, $\Delta E_{cphs}$ is
the excitation energy, and $\Delta k_{cphs}$ the corresponding excitation momentum. A
general energy eigenstate $\vert f\rangle$ with finite values for the numbers
$L_{c,\,-1/2}$ and/or $L_{s,\,-1/2}$, can be expressed as follows,
\begin{equation}
\vert f\rangle = \prod_{\alpha
=c,\,s}\frac{({\hat{S}}^{\dag}_{\alpha})^{L_{\alpha,\,-1/2}}}{
\sqrt{{\cal{C}}_{\alpha}}}\vert f.L\rangle \, . \label{IRREG}
\end{equation}
Here,
\begin{equation}
{\cal{C}}_{\alpha} = \delta_{L_{\alpha,\,-1/2},\,0} +
\prod_{l=1}^{L_{\alpha,\,-1/2}}l\,[\,L_{\alpha}+1-l\,] \, ; \hspace{1.0 cm}
L_{\alpha,\,-1/2}\leq L_{\alpha}=2S_{\alpha} \, , \label{Calpha}
\end{equation}
and the $\eta$-spin flip Yang holon ($\alpha =c$) and spin flip HL spinon ($\alpha =s$)
operators ${\hat{S}}^{\dag}_{\alpha}$ are the off-diagonal generators of the
corresponding $SU(2)$ algebras given in Eqs. (7) and (8), respectively, of Ref. \cite{I}.
These operators remain invariant under the electron - rotated-electron unitary
transformation and thus have the same expression in terms of electronic and
rotated-electron creation and annihilation operators. Moreover, in Eq. (\ref{IRREG})
$\vert f.L\rangle$ is the LWS that corresponds to the state $\vert f\rangle$. For a state
$\vert f\,;C\rangle$ belonging to a given CPHS the corresponding LWS is denoted by $\vert
f.L\,;C\rangle$. However, note that a non-LWS $\vert f\,;C\rangle$ and the corresponding
LWS $\vert f.L\,;C\rangle$ belong to different CPHSs, once they correspond to different
values of the numbers $L_{c,\,-1/2}$ and/or $L_{s,\,-1/2}$.

It is useful to reexpress the spectral-function expression (\ref{ABONj-HS}) in terms of
matrix elements between regular states only. The ground state is a LWS of both the
$\eta$-spin and spin $SU(2)$ algebras and thus has the following property,
\begin{equation}
{\hat{S}}_{\alpha}\,\vert GS\rangle = 0 \, ; \hspace{0.5cm} \alpha = c,\,s \, .
\label{GS=0}
\end{equation}

Let us introduce the operators ${\hat{\Theta}}_{{\cal{N}},\,j}^l$ and
${\hat{\Theta}}_{{\cal{N}},\,k}^{l}$ such that,
\begin{eqnarray}
\langle f.L;\,C\vert \prod_{\alpha =c,\,s}{1\over
\sqrt{{\cal{C}}_{\alpha}}}({\hat{S}}_{\alpha})^{L_{\alpha,\,-1/2}}\,
{\hat{O}}_{{\cal{N}},\,j}^{l} \vert GS\rangle & = & \Bigl[\prod_{\alpha =c,\,s}{1\over
\sqrt{{\cal{C}}_{\alpha}}}\Bigr]\,\langle f.L;\,C\vert
{\hat{\Theta}}_{{\cal{N}},\,j}^{l} \vert GS\rangle \, , \nonumber \\
\langle f.L;\,C\vert \prod_{\alpha =c,\,s}{1\over
\sqrt{{\cal{C}}_{\alpha}}}({\hat{S}}_{\alpha})^{L_{\alpha,\,-1/2}}\,
{\hat{O}}_{{\cal{N}}}^l (k) \vert GS\rangle & = & \Bigl[\prod_{\alpha =c,\,s}{1\over
\sqrt{{\cal{C}}_{\alpha}}}\Bigr]\,\langle f.L;\,C\vert {\hat{\Theta}}_{{\cal{N}},\,k}^{l}
\vert GS\rangle \, ; \hspace{0.5cm} l=\pm 1 \, . \label{SYMME}
\end{eqnarray}
By suitable use of Eq. (\ref{GS=0}), it is straightforward to show that the operators
${\hat{\Theta}}_{{\cal{N}},\,j}^{l}$ and ${\hat{\Theta}}_{{\cal{N}},\,k}^{l}$ are given
by the following commutators,
\begin{eqnarray}
{\hat{\Theta}}_{{\cal{N}},\,j}^{l} & = & \Bigl[\prod_{\alpha
=c,\,s}({\hat{S}}_{\alpha})^{L_{\alpha,\,-1/2}},\,
{\hat{O}}_{{\cal{N}},\,j}^{l}\Bigr] \, , \nonumber \\
{\hat{\Theta}}_{{\cal{N}},\,k}^{l} & = & \Bigl[\prod_{\alpha
=c,\,s}({\hat{S}}_{\alpha})^{L_{\alpha,\,-1/2}},\, {\hat{O}}_{{\cal{N}},\,k}^{l}\Bigr] \,
, \hspace{1.0cm} L_{c,\,-1/2}\hspace{0.2cm} {\rm and/or}\hspace{0.2cm}L_{s,\,-1/2}
> 0 \, ; \hspace{0.5cm} l=\pm 1 \, , \label{Th}
\end{eqnarray}
or by,
\begin{equation}
{\hat{\Theta}}_{{\cal{N}},\,j}^{l} = {\hat{O}}_{{\cal{N}},\,j}^{l} \, ; \hspace{1.0cm}
{\hat{\Theta}}_{{\cal{N}},\,k}^{l} = {\hat{O}}_{{\cal{N}},\,k}^{l} \, , \hspace{0.5cm}
L_{c,\,-1/2}=L_{s,\,-1/2}= 0 \, ; \hspace{0.5cm} l=\pm 1 \, . \label{OL-0}
\end{equation}
Thus,
\begin{eqnarray}
B_{{\cal{N}}}^{l} (k,\,\omega) & = & \sum_{\{\Delta
N_{\alpha\nu}\},\,\{L_{\alpha,\,-1/2}\}}\Bigl(\prod_{\alpha =c,\,s}{1\over
{\cal{C}}_{\alpha}}\Bigr)\,\sum_{f}\,N_a\vert\langle f.L;\,C\vert\,
{\hat{\Theta}}_{{\cal{N}},\,0}^{l} \vert GS\rangle\vert^2\nonumber
\\
& \times & \delta\Bigl(\omega - \Delta E_{cphs}\Bigr)\,\delta_{k,\,\Delta k_{cphs}} \, ;
\hspace{0.5cm} l=\pm 1 \, . \label{ABONj-L}
\end{eqnarray}
Note that when the operator ${\hat{\Theta}}_{{\cal{N}},\,0}^{l}$ is given by Eq.
(\ref{OL-0}) one has that $\vert f.L;\,C\rangle=\vert f;\,C\rangle$ in Eq.
(\ref{ABONj-L}). If the commutator $[\prod_{\alpha
=c,\,s}({\hat{S}}_{\alpha})^{L_{\alpha,\,-1/2}},\, {\hat{O}}_{{\cal{N}},\,j}^{l}]$ of Eq.
(\ref{Th}) vanishes, then the excitation generated by application of the corresponding
operator ${\hat{O}}_{{\cal{N}},\,j}^{l}$ onto the ground state has no overlap with the
excited energy eigenstate (\ref{IRREG}).

Similarly to ${\hat{O}}_{{\cal{N}},\,j}^{l}$, the corresponding operator
${\hat{\Theta}}_{{\cal{N}},\,j}^l$ can be written as a ${\cal{N}}$-electron operator. Let
the numbers ${\cal{N}}_{l_c,\,l_s}^{l}$ of Eq. (\ref{d}) refer to the operator
${\hat{O}}_{{\cal{N}},\,j}^{l}$ of the spectral-function expression of Eq. (\ref{ABONj}).
Then we call ${\bar{\cal{N}}}_{l_c,\,l_s}^{l}$ the corresponding electronic numbers of
the operator ${\hat{\Theta}}_{{\cal{N}},\,j}^{l}$ of the spectral-function expression of
Eq. (\ref{ABONj-L}). The values are such that ${\cal{N}}=\sum_{l_c,\,l_s=\pm
1}{\bar{\cal{N}}}_{l_c,\,l_s}^{l}=\sum_{l_c,\,l_s=\pm 1}{\cal{N}}_{l_c,\,l_s}^{l}$ and,
\begin{equation}
\sum_{l_s=\pm 1}{\bar{\cal{N}}}_{\pm 1,\,l_s}^{+1} = \sum_{l_s=\pm 1}{\cal{N}}_{\pm
1,\,l_s}^{l} \pm 2L_{c,\,-1/2}\, ; \hspace{0.5cm} \sum_{l_c,\,l_s=\pm 1}\delta_{l_c,
\,\pm l_s}\,{\bar{\cal{N}}}_{l_c,\,l_s}^{l} = \sum_{l_c,\,l_s=\pm 1}\delta_{l_c, \,\pm
l_s}\,{\cal{N}}_{l_c,\,l_s}^{l}\mp 2L_{s,\,-1/2} \, ; \hspace{0.25cm} l=\pm 1 \, .
\label{N-NL}
\end{equation}
These relations provide information about the numbers of electronic creation and
annihilation operators of the ${\cal{N}}$-electron operator
${\hat{\Theta}}_{{\cal{N}},\,j}^{l}$ expression relative to the corresponding numbers of
the ${\hat{O}}_{{\cal{N}},\,j}^{l}$ expression. While the number of electronic creation
(and annihilation) operators decreases (and increases) by $2L_{c,\,-1/2}$, the number of
electronic spin-down creation and spin-up annihilation (and spin-down annihilation and
spin-up creation) operators decreases (and increases) by $2L_{s,\,-1/2}$.

We note that the numbers ${\cal{N}}_{l_c,\,l_s}^{l}$ of Eq. (\ref{d}) for the operator
${\hat{O}}_{{\cal{N}},\,j}^{l}$ on the right-hand-side of Eq. (\ref{Th}) obey the sum
rules (\ref{NupdodllO}). Thus, following the relations of Eq. (\ref{N-NL}), the numbers
${\bar{\cal{N}}}_{l_c,\,l_s}^{l}$ of the corresponding operator
${\hat{\Theta}}_{{\cal{N}},\,j}^l$ are such that,
\begin{eqnarray}
\Delta L_c & = & 2\Delta S_c = -\Delta N +2L_{c,\,-1/2} = \sum_{l_c,\,l_s=\pm
1}(l_c)\,{\bar{\cal{N}}}_{l_c,\,l_s}^l \, ; \nonumber \\ \Delta L_s & = & 2\Delta S_s =
\Delta (N_{\uparrow}-N_{\downarrow}) +2L_{s,\,-1/2} = -\sum_{l_c,\,l_s=\pm
1}(l_c\,l_s)\,{\bar{\cal{N}}}_{l_c,\,l_s}^{l} \, . \label{Nupdodll}
\end{eqnarray}

The first relation of Eq. (\ref{NupdodllO}) just states that the difference in the number
of electronic creation and annihilation operators of the original ${\cal{N}}$-electron
operator ${\hat{O}}_{{\cal{N}},\,j}^{l}$ equals the value $\Delta N$ of the electron
number deviation generated by such operator. Similarly, the first relation of Eq.
(\ref{Nupdodll}) states that the difference in the number of rotated-electron
annihilation and creation operators of ${\hat{\Theta}}_{{\cal{N}},\,j}^{l}$ equals twice
the value $\Delta S_c=\Delta \eta$ of the $\eta$-spin value deviation generated by that
${\cal{N}}$-rotated-electron operator. Similar considerations apply to the second
relations of Eqs. (\ref{NupdodllO}) and (\ref{Nupdodll}).

We emphasize that all matrix elements of the general spectral-function expression
(\ref{ABONj-L}) refer to regular energy eigenstates. Indeed, by changing from the
spectral function representation (\ref{ABONj}) to (\ref{ABONj-L}) we have eliminated the
explicit presence of $-1/2$ Yang holons and $-1/2$ HL spinons. This was done by
evaluation of the contribution of these quantum objects to the $\cal{N}$-electron
spectral weight. Such a procedure corresponds to the computation of the commutator
$[\prod_{\alpha =c,\,s}({\hat{S}}_{\alpha})^{L_{\alpha,\,-1/2}},\,
{\hat{O}}_{{\cal{N}},\,j}^{l}]$ on the right-hand side of Eq. (\ref{Th}).

Our next step is the expression of the operator ${\hat{\Theta}}_{{\cal{N}},\,j}^{l}$ for
the general spectral-function expression (\ref{ABONj-L}) in terms of rotated-electron
creation and annihilation operators. Here we use the results of Refs. \cite{V,I,IV}
concerning the expression of the rotated electrons in terms of $\pm 1/2$ holons, $\pm
1/2$ spinons, and $c0$ pseudofermions. It is this direct relation that makes convenient
the rotated-electron expression for the $\cal{N}$-electron spectral functions. The
expression of the local ${\cal{N}}$-electron operator ${\hat{\Theta}}_{{\cal{N}},\,j}^l$
in terms of rotated-electron creation and annihilation operators is obtained by use of
the following relation,
\begin{equation}
{\hat{\Theta}}_{{\cal{N}},\,j}^l =
e^{\hat{S}}\,{\tilde{\Theta}}_{{\cal{N}}_0,\,j}^l\,e^{-\hat{S}}=
{\tilde{\Theta}}_{{\cal{N}}_0,\,j}^l +
\sum_{i=1}^{\infty}\sqrt{c^l_i}\,{\tilde{\Theta}}_{{\cal{N}}_i,\,j}^l \, ; \hspace{0.5cm}
j =1,2,...,N_a \, ; \hspace{0.5cm} l=\pm 1 \, . \label{ONjtil}
\end{equation}
Here $\hat{S}$ is the operator defined by Eqs. (21)-(23) of Ref. \cite{I} and
${\tilde{\Theta}}_{{\cal{N}}_0,\,j}^l$ has the same expression in terms of
rotated-electron creation and annihilation operators as
${\hat{\Theta}}_{{\cal{N}},\,j}^l$ in terms of electronic creation and annihilation
operators and thus ${\cal{N}}_0={\cal{N}}$. It is given by,
\begin{eqnarray}
{\tilde{\Theta}}_{{\cal{N}}_0,\,j}^l =
{\hat{V}}^{\dag}(U/t)\,{\hat{\Theta}}_{{\cal{N}},\,j}^l\,{\hat{V}}(U/t) =
e^{-\hat{S}}\,{\hat{\Theta}}_{{\cal{N}},\,j}^l\,e^{\hat{S}} \, ; \hspace{0.5cm} j
=1,2,...,N_a \, ; \hspace{0.5cm} l=\pm 1 \, . \nonumber
\end{eqnarray}

The operators ${\tilde{\Theta}}_{{\cal{N}}_i,\,j}^l$ on the right-hand side of Eq.
(\ref{ONjtil}) such that $i=1,2,...$ can be written as a product of ${\cal{N}}_i$
rotated-electron creation and annihilation operators and the value of the coefficient
$c^l_i$ is a function of $n$, $m$, and $U/t$ such that $c^l_i\rightarrow 0$ as
$U/t\rightarrow\infty$. For instance, for $i=1$ and $i=2$ we find,
\begin{equation}
\sqrt{c^l_1}\,{\tilde{\Theta}}_{{\cal{N}}_1,\,j}^l =
[{\hat{S}},\,{\tilde{\Theta}}_{{\cal{N}}_0,\,j}^l\,] \, ; \hspace{0.5cm} j =1,2,...,N_a
\, ; \hspace{0.5cm} l=\pm 1 \, , \label{ONjtil12}
\end{equation}
and
\begin{equation}
\sqrt{c^l_2}\,{\tilde{\Theta}}_{{\cal{N}}_2,\,j}^l = {1\over
2}\,[{\hat{S}},\,[{\hat{S}},\,{\tilde{\Theta}}_{{\cal{N}}_0,\,j}^l\,]\,] \, ;
\hspace{0.5cm} j =1,2,...,N_a \, ; \hspace{0.5cm} l=\pm 1 \, , \label{ONjtil12-B}
\end{equation}
respectively, and the $i>2$ operator terms are easily generated and involve similar
commutators. For simplicity, here we omit the longer expressions of the latter terms.

It is useful for the study of the spectral-function expressions to divide each CPHS
ensemble subspace in a set of well-defined subspaces. The number deviation $\Delta
N_{\alpha\nu}$ for the $\alpha\nu =c0,\,s1$ branches and the number $N_{\alpha\nu}=\Delta
N_{\alpha\nu}$ for the $\alpha\nu\neq c0,\,s1$ branches can be expressed in terms of
other related numbers as follows,

\begin{eqnarray}
\Delta N_{\alpha\nu} & = & \Delta N^F_{\alpha\nu}+\Delta N^{NF}_{\alpha\nu} \, ;
\hspace{0.5cm} \Delta N^F_{\alpha\nu}=\Delta N^F_{\alpha\nu ,\,+1}+\Delta N^F_{\alpha\nu
,\,-1} \, ; \hspace{0.5cm} 2\Delta J^F_{\alpha\nu}=\Delta N^F_{\alpha\nu ,\,+1}-\Delta
N^F_{\alpha\nu ,\,-1}
\, ; \nonumber \\
\Delta N^F_{\alpha\nu,\,\iota} & = & \Delta N^{0,F}_{\alpha\nu,\,\iota}+\iota\,Q^
0_{\alpha\nu}/2\pi \, ; \hspace{0.5cm} \Delta N^{0,F}_{\alpha\nu} = \Delta
N^{0,F}_{\alpha\nu ,\,+1}+\Delta N^{0,F}_{\alpha\nu ,\,-1} = \Delta N^F_{\alpha\nu} \, ;
\hspace{0.5cm} \alpha\nu = c0,\, s1 \hspace{0.25cm}
\iota = \pm 1 \, ; \nonumber \\
N_{\alpha\nu} & = & N_{\alpha\nu}^{F} + N_{\alpha\nu}^{NF} \, ; \hspace{0.5cm}
N^{F}_{\alpha\nu} = N^{F}_{\alpha\nu ,\,+1}+N^{F}_{\alpha\nu ,\,-1} \, ; \hspace{0.5cm}
2J^F_{\alpha\nu}=N^F_{\alpha\nu ,\,+1}-N^F_{\alpha\nu ,\,-1}\, ; \hspace{0.5cm}
\alpha\nu\neq c0,\, s1 \, , \label{numbers}
\end{eqnarray}
Here $\Delta N^F_{\alpha\nu ,\,\pm 1}$ is the deviation in the number of $\alpha\nu
=c0,\,s1$ pseudofermions at the right $(+1)$ and left right $(-1)$ {\it Fermi points},
$\Delta N^F_{\alpha\nu}$ and $\Delta J^F_{\alpha\nu}$ are the corresponding number and
current number deviations, respectively, and $\Delta N^{NF}_{\alpha\nu}$ gives the
deviation in the number of $\alpha\nu=c0,\,s1$ pseudofermions away from these points.
Moreover, $\Delta N^{0,F}_{\alpha\nu ,\,\pm 1}$ is the actual number of $\alpha\nu$
pseudofermions created or annihilated at the right $(+1)$ and left right $(-1)$ {\it
Fermi points} and $Q_{\alpha\nu}^0/2$ is the scattering-less phase shift on the
right-hand side of Eq. (\ref{Qcan1j}). For the $\alpha\nu\neq c0,\, s1$ branches,
$N_{\alpha\nu,\,\iota}^{F}$ is the number of $\alpha\nu$ pseudofermions with limiting
bare-momentum value $q=\iota\,q_{\alpha\nu}^0$ such that $\iota =\pm 1$,
$J^F_{\alpha\nu}$ is the corresponding current number, and $N_{\alpha\nu}^{NF}$ is the
number of $\alpha\nu$ pseudofermions whose bare-momentum values obey the inequality
$\vert q\vert<q_{\alpha\nu}^0$.

Let us also consider the number $N^{phNF}_{\alpha\nu}$ of finite-momentum and
finite-energy $\alpha\nu$ pseudofermion particle-hole processes (A), which refers to the
$\alpha\nu=c0,\,s1$ branches only \cite{V}.  $N^{phNF}_{\alpha\nu}$ is zero or a positive
integer such that $N^{phNF}_{\alpha\nu}=[{\cal{N}}_{\alpha\nu}-\vert\Delta
N^{NF}_{\alpha\nu}\vert]/2$. Here ${\cal{N}}_{\alpha\nu}$ gives the number of $\alpha\nu$
pseudofermion creation and annihilation operators involved in the expression of the
generators of the elementary processes (A).

The {\it J-CPHS ensemble subspaces} are the subspaces of a CPHS ensemble subspace spanned
by the excited energy eigenstates with the same values for the numbers $N^{phNF}_{c0}$,
$N^{phNF}_{s1}$, $\Delta N^F_{c0 ,\,+1}$, $\Delta N^F_{c0 ,\,-1}$, $\Delta N^F_{s1
,\,+1}$, $\Delta N^F_{s1 ,\,-1}$, and sets of numbers $\{N^F_{\alpha\nu ,\,+1}\}$ and
$\{N^F_{\alpha\nu ,\,-1}\}$ for the $\alpha\nu\neq c0,\,s1$ branches with finite
pseudofermion occupancy in the CPHS ensemble subspace.

Use of Eq. (\ref{ONjtil}) for $j=0$ in the general spectral-function expression
(\ref{ABONj-L}) leads to,
\begin{equation}
B_{{\cal{N}}}^{l} (k,\,\omega) = \sum_{i=0}^{\infty}c^l_i\sum_{\{\Delta
N_{\alpha\nu}\},\,\{L_{\alpha,\,-1/2}\}}\Bigl[\sum_{\{N^{phNF}_{\alpha\nu}\},\,\{\Delta
N^F_{\alpha\nu ,\,\iota}\},\, \{N^F_{\alpha\nu,\,\iota}\}}\,B^{l,i} (k,\,\omega)\Bigr] \,
; \hspace{0.5cm} c^l_0 = 1 \, , \hspace{0.5cm} l=\pm 1 \, , \label{ABONjl-CPHS}
\end{equation}
where the summations $\sum_{\{\Delta N_{\alpha\nu}\},\,\{L_{\alpha,\,-1/2}\}}$ and
$\sum_{\{N^{phNF}_{\alpha\nu}\},\,\{\Delta N^F_{\alpha\nu,\,\iota}\},\,\{N^F_{\alpha\nu
,\,\iota}\}}$ run over CPHS ensemble subspaces and the corresponding J-CPHS ensemble
subspaces of each of these spaces, respectively. The function $B^{l,i} (k,\,\omega)$ on
the right-hand side of Eq. (\ref{ABONjl-CPHS}) reads,
\begin{eqnarray}
B^{l,i} (k,\,\omega) & = & \Bigl(\prod_{\alpha =c,\,s}{1\over
{\cal{C}}_{\alpha}}\Bigr)\,\sum_{f}\, N_a\vert\langle f.L;\,JC\vert\,
{\tilde{\Theta}}_{{{\cal{N}}_i},\,0}^{l}\vert
GS\rangle\vert^2\nonumber \\
& \times & \delta\Bigl( \omega - l\Delta E_{j-cphs} \Bigr)\,\delta_{k,\,l\Delta
k_{j-cphs}} \, ; \hspace{0.5cm} l = \pm 1 \, ; \hspace{0.25cm} i=0,1,2,... \, ,
\label{ABONj+-cp}
\end{eqnarray}
where the summation $\sum_{f}$ runs over the excited energy eigenstates $\vert
f.L;\,JC\rangle$ which span each J-CPHS ensemble subspace. Thus, there is a function
$B^{l,i} (k,\,\omega)$ for each of these subspaces. (We recall that $\vert
f.L;\,JC\rangle$ is the LWS of a state $\vert f;\,JC\rangle$ related to it by the general
equation (\ref{IRREG}).)

Finally, let us use a notation for the number of spin-down and spin-up rotated-electron
creation and annihilation operators of the operator
${\tilde{\Theta}}_{{\cal{N}}_i,\,j}^{l}$ such that $i=0,1,2,...$ similar to that
associated with the numbers ${\bar{\cal{N}}}_{l_c,\,l_s}^{l}$ of Eq. (\ref{N-NL}). Thus,
we introduce the numbers,
\begin{equation}
{\cal{N}}_i =\sum_{l_c,\,l_s=\pm 1} {\bar{\cal{N}}}_{l_c,\,l_s}^{l,i} \, ; \hspace{0.5cm}
l= \pm 1 \, ; \hspace{0.5cm} i=0,1,2... \, , \label{di}
\end{equation}
which refer to the operator ${\tilde{\Theta}}_{{\cal{N}}_i,\,j}^{l}$. Here
${\bar{\cal{N}}}_{l_c,\,l_s}^{l,i}$ is the number of rotated-electron creation and
annihilation operators for $l_c=-1$ and $l_c=+1$, respectively, and with spin down and
spin up for $l_s=-1$ and $l_s=+1$, respectively. The operator
${\tilde{\Theta}}_{{\cal{N}}_0,\,j}^l$ of Eq. (\ref{ONjtil}) has the same four
rotated-electron numbers
$\{{\bar{\cal{N}}}_{l_c,\,l_s}^{0,l}\}=\{{\bar{\cal{N}}}_{l_c,\,l_s}^l\}$ as the
corresponding operator ${\hat{\Theta}}_{{\cal{N}},\,j}^{l}$ in terms of electrons.

\section{ROTATED-ELECTRON SUM RULES, SELECTION RULES, AND
ELEMENTARY PROCESSES IN TERMS OF PSEUDOFERMION OPERATORS} \label{SecIV}

In this section we provide sum rules and selection rules which for the PS arise from the
direct relation between rotated electrons and the holons, spinons, and pseudofermions.
Furthermore, we use such a relation to express the elementary rotated-electron processes
in terms of the pseudofermion creation and annihilation operators.

An important symmetry is that all six generators of the $\eta$-spin and spin $SU(2)$
algebras are invariant under the electron - rotated-electron unitary transformation
\cite{I}. Thus, the number of spin-projection $\sigma$ electrons equals the number of
spin-projection $\sigma$ rotated electrons. This also applies to the deviations $\Delta
S_c$ and $\Delta S_s$ in the $\eta$-spin and spin values, respectively, generated by
application onto the ground state of a $\cal{N}$-electron operator. This symmetry implies
that all $i=0,1,2,...$ operators ${\tilde{\Theta}}_{{\cal{N}}_i,\,j}^l$ on the right-hand
side of Eq. (\ref{ONjtil}) generate the same deviations $\Delta S_c$ and $\Delta S_s$, as
the operator ${\hat{\Theta}}_{{\cal{N}},\,j}^l$ on the left-hand side of the same
equation. It follows that the values ${\bar{\cal{N}}}_{l_c,\,l_s}^{l,i}$ for all these
$i=0,1,2,...$ operators with the same value of $l=\pm 1$ obey the following sum rules,
\begin{eqnarray}
\Delta L_c & = & 2\Delta S_c = -\Delta N +2L_{c,\,-1/2} = \sum_{l_c,\,l_s=\pm
1}(l_c)\,{\bar{\cal{N}}}_{l_c,\,l_s}^{l,i} \, ; \nonumber \\ \Delta L_s & = & 2\Delta S_s
= \Delta (N_{\uparrow}-N_{\downarrow}) +2L_{s,\,-1/2} = -\sum_{l_c,\,l_s=\pm
1}(l_c\,l_s)\,{\bar{\cal{N}}}_{l_c,\,l_s}^{l,i} \, ; \hspace{0.5cm} l = \pm 1 \, ;
\hspace{0.5cm} i=0,1,2,... \, . \label{Nupdodlli}
\end{eqnarray}
These rules provide useful information about the expression of all $i=1,2,...$ operators
${\tilde{\Theta}}_{{\cal{N}}_i,\,j}^{l}$. In addition to the rotated-electron creation
and annihilation operators of ${\tilde{\Theta}}_{{\cal{N}}_0,\,j}^{l}$, such an
expression includes pairs of rotated-electron creation and annihilation operators with
the same spin projection $\sigma$. Thus, such additional creation and annihilation
operators only generate rotated-electron particle-hole excitations and do not change the
net number of spin-projection $\sigma$ rotated electrons created or annihilated by
application of the operators ${\tilde{\Theta}}_{{\cal{N}}_i,\,j}^l$ onto the ground
state. The general situation refers to ${\cal{N}}$-electron operators that are not
invariant under the electron - rotated-electron unitary transformation. (The problem is
trivial for those that are invariant, once ${\tilde{\Theta}}_{{\cal{N}}_i,\,j}^l=0$ for
$i>0$ in that case.) The precise form of ${\tilde{\Theta}}_{{\cal{N}}_i,\,j}^l$ for
$i=1,2,...$ depends on the specific ${\cal{N}}$-electron operator under consideration.
However, a general property that follows from the relations (\ref{Nupdodlli}) is that for
increasing values of $i=1,2,...$ the operators ${\tilde{\Theta}}_{{\cal{N}}_i,\,j}^l$ are
constructed by adding to ${\tilde{\Theta}}_{{\cal{N}}_0,\,j}^l$ an increasing number of
{\it particle-hole} elementary spin-projection $\sigma$ rotated-electron pairs.

In equation (\ref{ONjtil}) the ${\cal{N}}$-electron operator
${\hat{\Theta}}_{{\cal{N}},\,j}^l$ is expressed in terms of rotated-electron creation and
annihilation operators. From the direct relation between the rotated electrons and the
holons, spinons, and pseudofermions it is straightforward to find useful selection rules.
Such rules refer to restrictions in the values of the number of $-1/2$ holons and thus of
$2\nu$-holon composite $c\nu$ pseudofermions generated by application onto the ground
state of each of the $i=0,1,2,...$ operators ${\tilde{\Theta}}_{{\cal{N}}_i,\,j}^l$ of
expression (\ref{ONjtil}). [The expression of these operators determines the value of the
$\cal{N}$-electron spectral function of Eq. (\ref{ABONjl-CPHS}), as confirmed by the form
of the related functions (\ref{ABONj+-cp}).] For the PS excited energy eigenstates which
have finite overlap with the $\cal{N}$-electron excitations, the values of the
$-1/2$-holon number, $M_{c,\,-1/2}$, and number of finite-energy and finite-momentum $c0$
pseudofermion particle-hole processes, $N^{phNF}_{c0}$, of the elementary processes (A)
\cite{V} are restricted to the following ranges,
\begin{eqnarray}
M_{c,\,-1/2} & = & L_{c,\,-1/2} + \sum_{\nu =1}^{\infty}\nu\,N_{c\nu}
=0,1,...,\sum_{l_s=\pm 1}{\bar{\cal{N}}}_{-1,\,l_s}^{l,i} \, ;
\nonumber \\
N^{phNF}_{c0} & = & 0,1,...,{\rm min}\left\{\sum_{l_s=\pm
1}{\bar{\cal{N}}}_{-1,\,l_s}^{l,i} ,\,\sum_{l_s=\pm
1}{\bar{\cal{N}}}_{+1,\,l_s}^{l,i}\right\} \, ; \hspace{0.5cm} i=0,1,2,... \, .
\label{src}
\end{eqnarray}
Here the numbers ${\bar{\cal{N}}}_{l_c,\,l_s}^{l,i}$ are those of Eq. (\ref{di}) for the
operator ${\tilde{\Theta}}_{{\cal{N}}_{i},\,j}^l$. The $i=0$ operator
${\tilde{\Theta}}_{{\cal{N}}_{0},\,j}^l$ has the same expression in terms of
rotated-electron creation and annihilation operators as the corresponding operator
${\hat{\Theta}}_{{\cal{N}},\,j}^l$ of Eq. (\ref{ONjtil}) in terms of electronic creation
and annihilation operators. Therefore, for $i=0$ the selection rules given in Eq.
(\ref{src}) read,
\begin{equation}
M_{c,\,-1/2} = L_{c,\,-1/2} + \sum_{\nu =1}^{\infty}\nu\,N_{c\nu} =0,1,...,\sum_{l_s=\pm
1}{\bar{\cal{N}}}_{-1,\,l_s}^{l} \, ; \hspace{0.5cm} N^{phNF}_{c0} = 0,1,...,{\rm
min}\left\{\sum_{l_s=\pm 1}{\bar{\cal{N}}}_{-1,\,l_s}^{l} ,\,\sum_{l_s=\pm
1}{\bar{\cal{N}}}_{+1,\,l_s}^{l}\right\} \, , \label{src0}
\end{equation}
where the numbers ${\bar{\cal{N}}}_{-1,\,l_s}^{l} ={\bar{\cal{N}}}_{-1,\,l_s}^{l,0}$ are
those of Eq. (\ref{N-NL}) specific to the operator ${\hat{\Theta}}_{{\cal{N}},\,j}^l$.

The first exact ground-state charge selection rule of Eq. (\ref{src}) concerning the
number of $-1/2$ holons, $M_{c,\,-1/2}$, is equivalent to the following selection rule
involving the number deviation $-\Delta M_c = \Delta M_s = \Delta N_{c0}$,
\begin{equation}
\sum_{l_c,\,l_s=\pm 1}(l_c)\,{\bar{\cal{N}}}_{l_c,\,l_s}^{l,i}\leq\Delta M_c\leq
{\cal{N}}_{i} \, ; \hspace{0.5cm} -{\cal{N}}_{i} \leq \Delta N_{c0} = \Delta M_s\leq
-\sum_{l_c,\,l_s=\pm 1}(l_c)\,{\bar{\cal{N}}}_{l_c,\,l_s}^{l,i}
 \, ; \hspace{0.5cm} i=0,1,2,... \, .
\label{McMs-range}
\end{equation}
Indeed, the combination of the inequalities (\ref{McMs-range}) with the relations
(27)-(29) of Ref. \cite{I} and that of Eq. (\ref{Nupdodlli}), readily confirms the
equivalence of the first selection rule given in Eq. (\ref{src}) and that of Eq.
(\ref{McMs-range}).

Moreover, the $-1/2$ spinon number deviation $\Delta M_{s,\,-1/2}$ is fully determined by
the value of the $-1/2$ holon number of the first selection rule of Eq. (\ref{src}) and
reads,
\begin{equation}
\Delta M_{s,\,-1/2} = L_{s,\,-1/2} + \Delta N_{s1} + \sum_{\nu =2}^{\infty}\nu\,N_{s\nu}
= \Delta N_{\downarrow} - M_{c,\,-1/2} \, . \label{Ms-}
\end{equation}

Equations (\ref{DNu}), (\ref{DNd}), (\ref{Nupdodlli}), and (\ref{Ms-}) define sum rules
obeyed by the values of the deviations in the quantum-object numbers and Eqs.
(\ref{srcsL}), (\ref{src}), and (\ref{McMs-range}) correspond to selection rules for the
permitted values of these deviations. Such sum rules and selection rules define the set
of CPHS ensemble subspaces which contain the excited energy eigenstates with finite
overlap with the $\cal{N}$-electron excitations under consideration.

While the above rules are exact, direct evaluation of the weights by the method
introduced in Sec. V and further developed in Ref. \cite{V} reveals that 94\% to 98\% of
the $\cal{N}$-electron weight corresponds to excited energy eigenstates with numbers in
the following range,
\begin{equation}
L_{s,\,-1/2}+\sum_{\nu =1}^{\infty}(\nu-1)\,N_{s\nu} =0,1,2,..., \sum_{l_c,\,l_s=\pm
1}\delta_{l_c,\,l_s}\, {\bar{\cal{N}}}_{l_c,\,l_s}^{l,i} \, ; \hspace{0.5cm} i=0,1,2,...
\, . \label{srs}
\end{equation}
For $i=0$ the relation (\ref{srs}) can be written as,
\begin{equation}
L_{s,\,-1/2}+\sum_{\nu =1}^{\infty}(\nu-1)\,N_{s\nu} =0,1,2,..., \sum_{l_c,\,l_s=\pm
1}\delta_{l_c,\,l_s}\, {\bar{\cal{N}}}_{l_c,\,l_s}^{l} \, , \label{srs0}
\end{equation}
where the numbers ${\bar{\cal{N}}}_{l_c,\,l_s}^{l} ={\bar{\cal{N}}}_{l_c,\,l_s}^{l,0}$
are those of Eq. (\ref{N-NL}) specific for the operator
${\hat{\Theta}}_{{\cal{N}},\,j}^l$ of Eq. (\ref{ONjtil}).

Local $-1/2$ holons (and $-1/2$ spinons) correspond to local $2\nu$-holon composite
$c\nu$ pseudofermions (and $2\nu$-spinon composite $s\nu$ pseudofermions). Local
$\alpha\nu$ pseudofermions are associated with the operators $f^{\dag
}_{x_j,\,\alpha\nu}$ and $f_{x_j,\,\alpha\nu}$ on the right-hand side of Eq. (34) of Ref.
\cite{V}. Let us denote the rotated-electron spin projections $\sigma =\uparrow
,\,\downarrow$ by $\sigma = -1/2,\,+1/2$, respectively, and consider the elementary
processes of the $\cal{N}$-electron excitations in terms of occupancy configurations of
local $\pm 1/2$ holons, $\pm 1/2$ spinons, and $c0$ pseudofermions \cite{IV}:
\vspace{0.3cm}

(i) To create one spin-projection $\sigma =\pm 1/2 $ rotated electron at the unoccupied
site $j$, we need to annihilate a local $+1/2$ holon and create a local $c0$
pseudofermion and a local $\pm 1/2$ spinon at the same site. Annihilation of a
spin-projection $\sigma =\pm 1/2$ rotated electron at a spin-projection $\sigma =\pm 1/2$
rotated-electron singly occupied site $j$, involves the opposite processes.\vspace{0.3cm}

(ii) To create one spin-projection $\sigma =\pm 1/2$ rotated electron at a
spin-projection $\sigma =\mp 1/2$ rotated-electron singly occupied site $j$, we need to
annihilate a local $\mp 1/2$ spinon and a local $c0$ pseudofermion and to create a local
$-1/2$ holon at such a site. Again, to annihilate a spin-projection $\sigma =\pm 1/2$
rotated electron at a rotated-electron doubly occupied site $j$, involves the opposite
processes.\vspace{0.3cm}

(iii) The creation of two rotated electrons of opposite spin projection onto the
unoccupied site $j$ involves the annihilation of a local $+1/2$ holon and the creation of
a local $-1/2$ holon at such a site. Annihilation of two rotated electrons of opposite
spin projection onto the doubly-occupied site $j$, involves the opposite
processes.\vspace{0.3cm}

(iv) The annihilation of one spin-projection $\sigma =\pm 1/2$ rotated electron and
creation of one spin-projection $\sigma =\mp 1/2$ rotated electron at the singly-occupied
site $j$, involves the annihilation of one local $\pm 1/2$ spinon and the creation of one
local $\mp 1/2$ spinon.\vspace{0.3cm}

Other processes can be expressed as suitable combinations of the above elementary
processes. The local rotated-electron operator terms which transfer spectral weight from
the ground state to each of the J-CPHS ensemble subspaces appearing in the state
summation of the spectral-function expression (\ref{ABONjl-CPHS}) have a specific and
uniquely defined form in terms of $c0$ pseudofermion and composite $\alpha\nu$
pseudofermion creation and annihilation operators. In order to find the pseudofermion
form of these operator terms it is crucial to take into account the initial ground-state
pseudofermion occupancies, given in Eqs. (C.24)-(C.25) of Ref. \cite{I}. (We recall that
the pseudoparticle-number values of the latter equations equal those of the corresponding
pseudofermion numbers.)

Before illustrating how the elementary processes (i)-(iv) are generated by the
pseudofermion creation and annihilation operators, it is convenient to provide some basic
rules for the use of the latter operators. Since following the use of the relations of
Eq. (\ref{SYMME}) all matrix elements are between the ground state and regular excited
states, in the processes considered below, the deviations in the numbers of Yang holons
($\alpha =c$) and HL spinons ($\alpha =s$) are such that $\Delta L_{\alpha} = \Delta
L_{\alpha \, ,+1/2}$. Some of these processes involve creation or annihilation of $+1/2$
Yang holons and/or $+1/2$ HL spinons. However, we recall that within the pseudofermion
representation, the $+1/2$ Yang holons and $+1/2$ HL spinons do not appear explicitly.
Such processes are taken into account by the deviations in the number of discrete
bare-momentum (and canonical-momentum) values and effective $\alpha\nu$ lattice sites of
the $c\nu\neq c0$ branches and/or $s\nu$ branches, respectively \cite{IV}. Given the
values of the corresponding pseudofermion number deviations, this is readily confirmed if
one compares the number (B.6) of Ref. \cite{I} of discrete bare-momentum values and of
effective lattice sites of the excited energy eigenstate and ground state CPHS ensemble
subspaces. Since $\Delta L_{\alpha} = \Delta L_{\alpha \, ,+1/2}$, note that following
Eqs. (B.6) and (B.7) of Ref. \cite{I}, the value of the number $N_{\alpha\nu}^*$ changes
when the value of the number $L_c = L_{c \, ,+1/2}$ of $+1/2$ Yang holons and/or $L_s =
L_{s \, ,+1/2}$ of $+1/2$ HL spinons also changes. Thus, creation and annihilation of
$+1/2$ Yang holons (and $+1/2$ HL spinons) are processes that are taken into account in
the definition of the effective $c\nu\neq c0$ pseudofermion lattices of the initial
ground state and excited energy eigenstates.

In the following we provide different examples of local rotated-electron operator
expressions in terms of pseudofermion creation and annihilation operators. For
simplicity, each of such pseudofermion expressions corresponds to the term of the local
rotated-electron operator which transfers spectral weight from the initial ground state
onto a single excitation J-CPHS ensemble subspace. Such a pseudofermion term includes a
coefficient factor $1/C_J$ whose value is well defined for each subspace. The full local
rotated-electron operator term which transfers spectral weight from the ground state to a
given J-CPHS ensemble subspace is the product of that studied here by another
pseudofermion operator term given in the ensuing section. (The general form of $1/C_J$ is
also given in that section.) The latter operator transfers from the ground state to the
J-CPHS ensemble subspace the part of the spectral weight which corresponds to the
processes (B) and (C), whereas the pseudofermion terms studied here transfer the part of
that weight associated with the processes (A).

A $i=0$ local rotated-electron operator ${\tilde{\Theta}}_{{\cal{N}}_0,\,j}^l$ always has
one or a few {\it dominant} CPHS ensemble subspaces which correspond to the whole
spectral weight transferred from the ground state in well defined limits. For
one-electron problems such that ${\cal{N}}_0 ={\cal{N}}=1$ this refers to the limits
where the spectral-weight distribution is $\delta$-function like, as in Eq. (77) of Ref.
\cite{V}. For ${\cal{N}}_0 ={\cal{N}}=2$, to the limits where such a distribution can be
expressed as a simple integral whose integrand is a $\delta$ function, as in Eq. (78) of
the same reference. In the general ${\cal{N}}_0 ={\cal{N}}>2$ case, to the limits where
the spectral-weight distribution can be written as an integral whose integrand is a
product of $[{\cal{N}}-1]$ $\delta$ functions. For instance, for ${\cal{N}}=1$ this
occurs for $U/t\rightarrow 0$. For the one-electron problem the amount of spectral weight
transferred from the ground state to the set of J-CPHS ensemble subspaces contained in
the dominant CPHS ensemble subspaces is weakly dependent on $U/t$: while for $U/t<<1$ it
corresponds to the whole spectral weight transferred from the ground state by the local
rotated-electron operator, for $U/t>>1$ it corresponds typically to more than 0.94\% of
that weight. (Comparison of the amount of transferred weight for $U/t<<1$ with that
obtained by use of the methods of Ref. \cite{Penc97} for $U/t>>1$ confirms such a weak
$U/t$ dependence.) For $U/t$ finite there arise an infinite number of pseudofermion
terms, each corresponding to a J-CPHS ensemble subspace compatible with the local
rotated-electron operator. Another example is the ${\cal{N}}=2$ charge dynamical
structure factor, where there are different dominant CPHS ensemble subspaces for
$U/t\rightarrow 0$ and $U/t\rightarrow\infty$, respectively. In this case the amount of
spectral weight transferred from the ground state to the set of J-CPHS ensemble subspaces
contained in the dominant CPHS ensemble subspaces is a decreasing (and increasing)
function of $U/t$ for the $U/t\rightarrow 0$ (and $U/t\rightarrow\infty$) dominant
subspaces and vanishes as $U/t\rightarrow\infty$ (and $U/t\rightarrow 0$). Again, for
intermediate finite values of $U/t$ there arise an infinite number of pseudofermion
terms, each corresponding to a J-CPHS ensemble subspace compatible with the local
rotated-electron operator. However, for all $i=0$ local rotated-electron operators
${\tilde{\Theta}}_{{\cal{N}}_0,\,j}^l$ the pseudofermion terms associated with the
dominant subspaces together with a small number of other terms correspond to more than
99\% of the spectral weight. It follows that in applications of the pseudofermion
dynamical theory introduced here and in Ref. \cite{V} only a finite number of
pseudofermion terms should be considered.

The $\alpha\nu =c0,\,s1$ pseudofermion number deviations and $\alpha\nu\neq c0,\,s1$
pseudofermion numbers are related to the rotated-electron number deviations by Eqs.
(\ref{DNu}) and (\ref{DNd}). Given the values of the $\alpha\nu =c0,\,s1$ pseudofermion
number deviations and $\alpha\nu\neq c0,\,s1$ pseudofermion numbers of the specific
J-CPHS ensemble subspace under consideration, the expression of the local
rotated-electron operator in terms of pseudofermion creation and annihilation operators
is always uniquely defined. Let us start by providing some of the simplest pseudofermion
operator terms of local rotated-electron operators. For local one- and
two-rotated-electron operators these operator terms involve in general $\alpha\nu$
pseudofermion creation and/or annihilation operators belonging to branches such that $\nu
<2$. The case of other terms associated with excitation J-CPHS ensemble subspaces
generated from the ground state by processes involving creation of composite $\alpha\nu$
pseudofermions for $\nu>1$ is discussed later.

In the following expressions the $\alpha\nu$ effective lattice integer site index $j'$ is
such that $j'=1,2,...,N_{\alpha\nu}^*$ \cite{IV}. An important property is that an
operator whose expression in terms of rotated-electron operators is local at $x_j = ja
=j$ can be written as a product of local $\alpha\nu$ pseudofermion operators at
$x_{j'}\approx x_j$ where $x_{j'} = j'a_{\alpha\nu}^0$. Here and in all expressions given
below $j'$ is defined for the $\alpha\nu\neq c0$ branches as the closest integer number
to $j n_{\alpha\nu}^*$, whereas $j' =j$ for $\alpha\nu =c0$. We note that for the former
branches the site $j'$ occupied by one $\alpha\nu$ pseudofermion corresponds to $2\nu$
sites of the rotated-electron lattice. Thus, $\vert x_{j'}-x_j\vert$ is always smaller
than the very small intrinsic uncertainty which corresponds to the $2\nu$
rotated-electron lattice sites occupied by the local $\alpha\nu$ pseudofermion. Moreover,
we emphasize that the rotated-electron lattice site $j$ associated with the effective
$\alpha\nu\neq c0$ lattice site $j'\approx j n_{\alpha\nu}^*$ defined above always
belongs to the domain of $2\nu$ rotated-electron lattice sites of $j'$. Here and below we
use the equality $j'=j n_{\alpha\nu}^*$ to denote the integer number $j'$ defined above.
Thus, the site $j'$ is such that $j'=j$ for $\alpha\nu =c0$ operators, $j' =j
n_{\uparrow}$ for $\alpha\nu =s1$ operators, $j'=j [1-n]$ for $c\nu \neq c0$ operators
when $n<1$, and $j' =j [n_{\uparrow}-n_{\downarrow}]$ for $s\nu\neq s1$ operators when
$m>0$. The following local rotated-electron operator expressions in terms of
pseudofermions, whose coefficient $C_J$ is different for each operator, refer to the
elementary processes (A) subspace:\vspace{0.3cm}

(i) One of the simplest processes for creation of one spin-down rotated electron at the
unoccupied site $j$ involves the creation of a local $c0$ pseudofermion with the operator
$f^{\dag }_{x_j,\,c0}$ and of a local $s1$ pseudofermion with the operator $f^{\dag
}_{x_{j'},\,s1}$ such that $j'=jn_{\uparrow}$,
\begin{equation}
{\tilde{c}}_{j,\,\downarrow}^{\dag}\,(1-{\tilde{n}}_{j,\,\uparrow}) = {1\over
C_{J}}\,f^{\dag }_{x_{j'},\,s1}\,f^{\dag }_{x_j,\,c0} \, , \label{cjpm0}
\end{equation}
and thus ${\tilde{c}}_{j,\,\downarrow}\,(1-{\tilde{n}}_{j,\,\uparrow}) = {1\over
C_{J}}\,f_{x_j,\,c0}\,f_{x_{j'},\,s1}$ refers to annihilation of one spin-down rotated
electron at the singly-occupied site $j$. In Eq. (\ref{cjpm0}),
${\tilde{n}}_{j,\,\sigma}={\tilde{c}}_{j,\,\sigma}^{\dag} \,{\tilde{c}}_{j,\,\sigma}$ is
the local spin-projection $\sigma$ rotated-electron density operator.

To create one spin-up rotated electron at the empty site $j$, a simple process
corresponds to create a local $c0$ pseudofermion with the operator $f^{\dag
}_{x_j,\,c0}$,
\begin{equation}
{\tilde{c}}_{j,\,\uparrow}^{\dag}\,(1-{\tilde{n}}_{j,\,\downarrow}) = {1\over
C_{J}}\,f^{\dag }_{x_j,\,c0} \, , \label{cjpm0up}
\end{equation}
and ${\tilde{c}}_{j,\,\uparrow}\,(1-{\tilde{n}}_{j,\,\downarrow}) = {1\over
C_{J}}\,f_{x_j,\,c0}$ to annihilation of one spin-up rotated electron at the
singly-occupied site $j$. Such processes also involve creation and annihilation,
respectively, of an empty site in the effective $s1$ lattice. When the initial ground
state belongs to a $m=0$ CPHS ensemble subspace, there is for the former process a single
$s1$ pseudofermion hole in the excited state, which corresponds to the created site.

We note that the processes of the first expressions of Eqs. (\ref{cjpm0}) and
(\ref{cjpm0up}) also involve the annihilation of a $+1/2$ Yang holon, whereas the
processes of the second expressions of the same equations involve the creation of a
$+1/2$ Yang holon. Similarly, the processes of the first and second expressions of Eq.
(\ref{cjpm0up}) involve the creation and annihilation, respectively, of a $+1/2$ HL
spinon. In the remaining cases considered below we do not specify the elementary
processes of creation or annihilation of $+1/2$ Yang holons and $+1/2$ HL spinons, which
are taken into account implicitly by the pseudofermion description, as discussed above.
\vspace{0.3cm}

(ii) One of the simplest processes associated with the creation of one spin-up rotated
electron at a spin-down rotated-electron singly occupied site $j$ involves the
annihilation a local $c0$ pseudofermion with the operator $f_{x_j,\,c0}$ and of a local
$s1$ pseudofermion with the operator $f_{x_{j'},\,s1}$ and the creation of a local $c1$
pseudofermion with the operator $f^{\dag }_{x_{j''},\,c1}$ such that $j'=jn_{\uparrow}$
and $j''=j[1-n]$, respectively,
\begin{equation}
{\tilde{c}}_{j,\,\uparrow}^{\dag}\,{\tilde{n}}_{j,\,\downarrow} = {1\over
C_{J}}\,f_{x_{j''},\,c1}^{\dag}\,f_{x_j,\,c0}\,f_{x_{j'},\,s1} \, . \label{cjpm1}
\end{equation}
Then ${\tilde{c}}_{j,\,\uparrow}\,{\tilde{n}}_{j,\,\downarrow} = {1\over
C_{J}}\,f_{x_{j'},\,s1}^{\dag}\,f_{x_j,\,c0}^{\dag}\,f_{x_{j''},\,c1}$ refers to
annihilation of one spin-up rotated electron at a doubly occupied site $j$. Moreover, to
create one spin-down rotated electron at a spin-up rotated-electron singly occupied site
$j$, a simple process corresponds to annihilate a local $c0$ pseudofermion with the
operator $f_{x_j,\,c0}$ and to create a local $c1$ pseudofermion with the operator
$f^{\dag }_{x_{j'},\,c1}$ such that $j'=j[1-n]$,
\begin{equation}
{\tilde{c}}_{j,\,\downarrow}^{\dag}\,{\tilde{n}}_{j,\,\uparrow} = {1\over
C_{J}}\,f_{x_{j'},\,c1}^{\dag}\,f_{x_j,\,c0} \, . \label{cjpm1D}
\end{equation}
In this case ${\tilde{c}}_{j,\,\downarrow}\,{\tilde{n}}_{j,\,\uparrow} = {1\over
C_{J}}\,f_{x_j,\,c0}^{\dag}\,f_{x_{j'},\,c1}$ corresponds to annihilation of one
spin-down rotated electron at a doubly occupied site $j$. \vspace{0.3cm}

(iii) A simple process involved in the creation of two rotated electrons of opposite spin
projection onto the empty site $j$ corresponds to creation of a local $c1$ pseudofermion
with the operator $f^{\dag }_{x_{j'},\,c1}$ such that $j'=j[1-n]$,
\begin{equation}
{\tilde{c}}_{j,\,\downarrow}^{\dag}\,{\tilde{c}}_{j,\,\uparrow}^{\dag} = {1\over
C_{J}}\,f_{x_{j'},\,c1}^{\dag} \, . \label{cjpm2}
\end{equation}
It follows that ${\tilde{c}}_{j,\,\uparrow}\,{\tilde{c}}_{j,\,\downarrow} = {1\over
C_{J}}\,f_{x_{j'},\,c1}$ refers to annihilation of two rotated electrons of opposite spin
projection onto a doubly-occupied site $j$. This involves annihilation of a local $c1$
pseudofermion with the operator $f_{x_{j'},\,c1}$ such that $j'=j[1-n]$. \vspace{0.3cm}

(iv) One of the simplest processes associated with the annihilation of one spin-up
rotated electron and creation of one spin-down rotated electron at the singly-occupied
site $j$, involves the creation of a local $s1$ pseudofermion with the operator $f^{\dag
}_{x_{j'},\,s1}$ such that $j'=jn_{\uparrow}$,
\begin{equation}
{\tilde{c}}_{j,\,\downarrow}^{\dag}\,{\tilde{c}}_{j,\,\uparrow} = {1\over C_{J}}\,f^{\dag
}_{x_{j'},\,s1} \, . \label{cjpmsf}
\end{equation}
Then ${\tilde{c}}_{j,\,\downarrow}^{\dag}\,{\tilde{c}}_{j,\,\uparrow} = {1\over
C_{J}}\,f_{x_{j'},\,s1}$ corresponds to annihilation of one spin-down rotated electron
and creation of one spin-up rotated electron at the singly-occupied site $j$. This
involves the annihilation of a local $s1$ pseudofermion with the operator
$f_{x_{j'},\,s1}$ such that $j'=jn_{\uparrow}$. \vspace{0.3cm}

For local rotated-electron operators generating more complex processes involving creation
or annihilation of several rotated electrons, the creation and annihilation of local $c0$
pseudofermions is always associated with creation and annihilation of rotated-electron
singly occupied sites, respectively. Since the local $c0$ pseudofermions and $c0$
pseudofermion holes occupy the same sites $j_l$ and $j_h$ as the rotated-electron singly
occupied sites and rotated-electron doubly-occupied and unoccupied sites, respectively,
there is a one-to-one correspondence between the rotated-electron and $c0$ pseudofermion
algebras. Here we have used the site notation of Ref. \cite{IV}.

However, once the composite local $\alpha\nu\neq c0$ pseudofermions have internal
structure that involves $2\nu$ rotated-electron sites with different index $j$, the
operational relation of rotated electrons to such composite quantum objects is more
involved \cite{IV}. This justifies why the expressions of the local rotated-electron
operators in terms of creation and annihilation pseudofermion operators involve a
superposition of different pseudofermion expressions, corresponding to the set of
compatible J-CPHS ensemble subspaces. Nevertheless, creation onto the ground state of a
$c\nu$ pseudofermion (and $s\nu$ pseudofermion) always involves $\nu$ rotated-electron
doubly occupied sites and $\nu$ unoccupied sites (and $\nu$ spin-down rotated-electron
singly occupied sites $\nu$ spin-up rotated-electron singly occupied sites). In general,
the excited-energy-eigenstate $\nu$ rotated-electron unoccupied sites (and $\nu$ spin-up
rotated-electron singly occupied sites) of a created $c\nu$ pseudofermion (and $s\nu$
pseudofermion) are generated by annihilating an equal number of $+1/2$ Yang holons (and
$+1/2$ HL spinons) of the initial ground state.

For the creation of a local $c\nu$ pseudofermion, each of the $\nu$ new created
rotated-electron doubly occupied sites can result from creation of a rotated-electron
pair onto an unoccupied site or of a rotated electron onto a singly-occupied site. The
latter case involves always one of the elementary processes associated with the
pseudofermion terms given just after Eqs. (\ref{cjpm0}) and (\ref{cjpm0up}). On the other
hand, for the creation of a local $s\nu$ pseudofermion such that $\nu>1$, the $\nu$
involved spin-down rotated-electron singly occupied sites can result from creation of
spin-down rotated electrons onto unoccupied sites or from recombination of pre-existing
ground-state $s1$ pseudofermions, as further discussed below.

For instance, let us consider two J-CPHS ensemble subspaces contained in different CPHS
ensemble subspaces which except for the occupancies of the $c1$ and $c2$ branches have
the same pseudofermion numbers. For such branches, one has $\{N_{c1}=2,\,N_{c2}=0\}$ for
the J-CPHS ensemble subspace (I) and $\{N_{c1}=0,\,N_{c2}=1\}$ for the J-CPHS ensemble
subspace (II). Let us consider that the local rotated-electron operator behind the
transitions to both subspaces is the same and involves creation of two rotated electrons
of spin projections $\sigma =\uparrow$ and $\sigma =\downarrow$ onto the spin-down singly
occupied site $j$ and spin-up singly occupied site $j +1$, respectively. For the subspace
(I), this corresponds simply to the process of Eq. (\ref{cjpm1}) for the site $j$ and the
process of Eq. (\ref{cjpm1D}) for the site $j+1$. For the subspace (II), in order to
create two rotated electrons of spin projections $\sigma =\uparrow$ and $\sigma
=\downarrow$ onto the spin-down singly occupied site $j$ and spin-up singly occupied site
$j +1$, respectively, we need to annihilate two local $c0$ pseudofermions with the
operators $f_{x_j,\,c0}$ and $f_{x_{j+1},\,c0}$ and a local $s1$ pseudofermion with the
operator $f_{x_{j'},\,s1}$ such that $j'=jn_{\uparrow}$ and to create a local $c2$
pseudofermion with the operator $f^{\dag }_{x_{j''},\,c2}$ such that $j''=j[1-n]$,
\begin{equation}
{\tilde{c}}_{j,\,\uparrow}^{\dag}\, {\tilde{c}}_{j,\,\downarrow}^{\dag}\,
{\tilde{n}}_{j,\,\downarrow}\, {\tilde{n}}_{j,\,\uparrow} = {1\over
C_{J}}\,f_{x_{j''},\,c2}^{\dag}\,f_{x_j,\,c0}\,f_{x_{j'},\,s1}\,\,f_{x_{j+1},\,c0} \, .
\label{cjpm2+}
\end{equation}
It should be mentioned that in spite of the annihilation of one $s1$ pseudofermion, this
process does not involve the corresponding creation of a $s1$ pseudofermion hole. Indeed,
it involves the annihilation of the site $j'=jn_{\uparrow}$ in the effective $s1$
lattice. Thus, when the initial ground state belongs to a $m=0$ CPHS ensemble subspace,
in spite of the annihilation of the $s1$ pseudofermion the excited state corresponds to a
fully occupied $s1$ band, as the initial ground state.

A similar process gives rise to creation of a local $s2$ pseudofermion provided that
creation of the two rotated-electron doubly-occupied sites is replaced by creation of two
spin-down rotated electron singly occupied sites. However, in this case there is the
possibility that one (or both) the spin-down spinons needed for creation of the local
$s2$ pseudofermion is (or are) generated from annihilation of one (or two) ground-state
$s1$ pseudofermion(s). Such processes can {\it dress} any rotated-electron process and
are behind the occurrence of an infinite number of compatible J-CPHS ensemble subspaces
for each local rotated-electron operator. These non-dominant pseudofermion processes do
not obey the relation (\ref{srs0}) (which is not an exact rotated-electron selection
rule) and for all finite values of $U/t$ amount to less than 6\% of the rotated-electron
spectral weight. For instance, in order to create one spin-down rotated electron at the
empty site $j$, in addition to the pseudofermion process (\ref{cjpm0}) there is for
instance a process corresponding to the creation of a $c0$ pseudofermion with the
operator $f^{\dag }_{x_j,\,c0}$ and of a $s2$ pseudofermion with the operator $f^{\dag
}_{x_{j''},\,s2}$ such that $j''=j[n_{\uparrow}-n_{\downarrow}]$ and to the annihilation
of a $s1$ pseudofermion with the operator $f^{\dag }_{x_{j'},\,s1}$ such that
$j'=jn_{\uparrow}$,
\begin{equation}
{\tilde{c}}_{j,\,\downarrow}^{\dag}\,(1-{\tilde{n}}_{j,\,\uparrow}) = {1\over
C_{J}}\,f^{\dag }_{x_{j''},\,s2}\,f_{x_{j'},\,s1}\,f^{\dag }_{x_j,\,c0} \, .
\label{cjpm0-2}
\end{equation}
We emphasize that the amount of spectral weight transferred from the ground state by the
operator (\ref{cjpm0}) is much larger than that transferred by the operator
(\ref{cjpm0-2}). Indeed, the J-CPHS ensemble subspace associated with the expression
(\ref{cjpm0}) belongs the dominant CPHS ensemble subspace of the local rotated-electron
operator, whereas the J-CPHS ensemble subspace corresponding the expression
(\ref{cjpm0-2}) does not. Note that in the present case the two competing J-CPHS ensemble
subspaces can have the same pseudofermion deviation numbers and values except for the
occupancies of the $s1$ and $s2$ branches. Thus, the processes generated by the operators
(\ref{cjpm0}) and (\ref{cjpm0-2}) correspond to J-CPHS ensemble subspaces belonging to
different CPHS ensemble subspaces such that $\{\Delta N_{s1}=1,\,N_{s2}=0\}$ and
$\{\Delta N_{s1}=-1,\,N_{s2}=2\}$, respectively. Similar {\it dressing processes}
involving creation of $s\nu$ pseudofermions belonging to $\nu>1$ branches by annihilation
of one to $\nu$ ground-state $s1$ pseudofermions can occur for all rotated-electron
processes but correspond to very fast decreasing values of the amount of spectral weight
transferred from the ground state for increasing number of pseudofermion processes.
Moreover, we recall that the subspace summation on the right-hand side of the
spectral-function expression (\ref{ABONjl-CPHS}) is limited to the compatible CPHS
ensemble subspaces: their pseudofermion number deviations and numbers obey the sum rules
(\ref{DNu}), (\ref{DNd}), (\ref{Nupdodlli}), and (\ref{Ms-}) and selection rules
(\ref{srcsL}), (\ref{src}), and (\ref{McMs-range}).

We could present here other pseudofermion terms of increasing complexity, corresponding
to the local rotated-electron operators considered above. However, the amount of spectral
weight transferred from the ground state by the pseudofermion operator terms describing
the above-mentioned dressing processes involving creation of $s\nu$ pseudofermions such
that $\nu>1$ by annihilation of an increasing number of ground-state $s1$ pseudofermions
decreases very rapidly for increasing values of $\nu$. Also the spectral weight
transferred from the ground state by the operator terms with increasing value for the
index $i$ of the expression (\ref{ONjtil}) of the general operator
${\hat{\Theta}}_{{\cal{N}},\,j}^l$ decreases very rapidly. For instance, for the
one-electron spectral weight the contributions from dressing $s\nu$ pseudofermion
processes for $\nu>2$ and the terms of index $i>1$ of the expression (\ref{ONjtil}) for
${\cal{N}} =1$ are typically beyond numerical measurability. Therefore, as far as
numerical measurability is concerned, only a few pseudofermion terms contribute to the
actual electronic spectral-weight distributions \cite{V}.

\section{THE SPECTRAL FUNCTION AS A CONVOLUTION OF
PSEUDOFERMION SPECTRAL FUNCTIONS}\label{SecV}

In this section we express the spectral functions (\ref{ABONj+-cp}) as a convolution of
pseudofermion, independent $-1/2$ holon, and independent $-1/2$ spinon spectral
functions. The excited energy eigenstates appearing on the right-hand side of Eq.
(\ref{ABONj+-cp}) can be written as the following pseudofermion Slater determinant,
\begin{equation}
\vert f.L,\,C\rangle = \prod_{\alpha\nu}\,F_{f,\,\alpha\nu}^{\dag}\vert 0\rangle \, ;
\hspace{1cm} F_{f,\,\alpha\nu}^{\dag} = \prod_{{{\bar{q}}_j}=
-q_{\alpha\nu}^0}^{+q_{\alpha\nu}^0}\,\Bigl[ {\cal{N}}_{\alpha\nu}({\bar{q}}_j)\,f^{\dag
}_{{\bar{q}},\,\alpha\nu} + [1 - {\cal{N}}_{\alpha\nu}({\bar{q}}_j)]\Bigr] \, .
\label{excSD}
\end{equation}
Here and in other expressions below $\vert 0\rangle$ is the pseudofermion vacuum such
that $f_{{\bar{q}_j},\,\alpha\nu}\vert 0\rangle =0$ for all $\alpha\nu$ branches and
canonical-momentum values. In turn, according to Eqs. (C.24) and (C.25) of Ref. \cite{I},
the ground state corresponds to a canonical-momentum densely packed occupancy for the
$c0$ and $s1$ pseudofermion bands and the Slater determinant has the following simpler
form,
\begin{eqnarray}
\vert GS\rangle & = & \prod_{\alpha\nu = c0,\,s1}\, F_{GS,\,\alpha\nu}^{\dag}\vert
0\rangle \, ; \hspace{0.5cm} F_{GS,\,\alpha\nu}^{\dag} = \prod_{{{\bar{q}}_j}=
-q_{F\alpha\nu}^0}^{+q_{F\alpha\nu}^0}\,f^{\dag }_{{\bar{q}_j},\,\alpha\nu} \, ;
\hspace{0.25cm} \alpha\nu = c0,\,s1
\, ; \nonumber \\
F_{-GS,\,\alpha\nu}^{\dag} & = & \prod_{{{\bar{q}}_j}=
q_{F\alpha\nu,\,-1}}^{q_{F\alpha\nu,\,+1}}\,f^{\dag }_{{\bar{q}_j},\,\alpha\nu} \, ;
\hspace{0.5cm} F_{J-GS,\,\alpha\nu}^{\dag} = \prod_{{{\bar{q}}_j}=
{\bar{q}}_{F\alpha\nu,\,-1}}^{{\bar{q}}_{F\alpha\nu,\,+1}}\,f^{\dag
}_{{\bar{q}_j},\,\alpha\nu} \, ; \hspace{0.25cm} \alpha\nu = c0,\,s1 \, . \label{GS-SD}
\end{eqnarray}
The generators $F_{-GS,\,\alpha\nu}^{\dag}$ and $F_{J-GS,\,\alpha\nu}^{\dag}$ given here
correspond to densely packed distributions introduced below and the discrete
canonical-momentum values of the pseudofermion operators $f^{\dag
}_{{\bar{q}_j},\,\alpha\nu}$ of their expressions are those of the CPHS ensemble subspace
which the ground state $\vert GS\rangle$ and the excited state $\vert f.L;\,C\rangle$ of
Eq. (\ref{excSD}) belong to, respectively. The {\it Fermi points} appearing in the
products of their expressions of Eq. (\ref{GS-SD}) read $q_{F\alpha\nu,\,\pm 1}=\pm
q_{F\alpha\nu}^0\pm [2\pi/L]\Delta N^{0,F}_{\alpha\nu ,\,\pm 1}$ and
${\bar{q}}_{F\alpha\nu,\,\pm 1}=\pm q_{F\alpha\nu}^0\pm [2\pi/L][\Delta N^F_{\alpha\nu
,\,\pm 1}\pm Q^{\Phi}_{\alpha\nu} (\pm q_{F\alpha\nu}^0)/2\pi]$, respectively, where the
deviation numbers $\Delta N^{0,F}_{\alpha\nu ,\,\pm 1}$ and $\Delta N^F_{\alpha\nu ,\,\pm
1}$ are those of Eq. (\ref{numbers}).

The excited-energy-eigenstate canonical-momentum distribution function
${\cal{N}}_{\alpha\nu}({\bar{q}}_j)$ on the right-hand side of Eq. (\ref{excSD}) can be
written as,
\begin{eqnarray}
{\cal{N}}_{\alpha\nu}({\bar{q}}_j) & = & {\cal{N}}_{\alpha\nu}^{ph}({\bar{q}}_j) + \Delta
{\cal{N}}_{\alpha\nu}^{NF}({\bar{q}}_j) \, ; \hspace{0.5cm}
{\cal{N}}_{\alpha\nu}^{ph}({\bar{q}}_j) = {\cal{N}}_{\alpha\nu}^{-0}({\bar{q}}_j) +
\Delta {\cal{N}}_{\alpha\nu}^{phF}({\bar{q}}_j) \, ; \hspace{0.25cm}
\alpha\nu = c0,\,s1 \nonumber \\
{\cal{N}}_{\alpha\nu}^{-0}({\bar{q}}_j) & = & {\cal{N}}_{\alpha\nu}^{0}({\bar{q}}_j) +
\Delta {\cal{N}}_{\alpha\nu}^{F}({\bar{q}}_j) \, ; \hspace{0.25cm} \alpha\nu = c0,\,s1 \,
; \hspace{0.5cm} {\cal{N}}_{\alpha\nu}({\bar{q}}_j) = \Delta
{\cal{N}}_{\alpha\nu}^{NF}({\bar{q}}_j) +\Delta {\cal{N}}_{\alpha\nu}^{F}({\bar{q}}_j) \,
; \hspace{0.25cm} \alpha\nu \neq c0,\,s1 \, . \label{NanphTh}
\end{eqnarray}
Here ${\cal{N}}_{\alpha\nu}^{-0}({\bar{q}}_j)$ and $N_{\alpha\nu}^{-0} (q_j)$, such that
${\cal{N}}_{\alpha\nu}^{-0}({\bar{q}}_j) =N_{\alpha\nu}^{-0} (q_j)$, correspond to the
excited densely packed distributions $\prod_{\alpha\nu = c0,\,s1}\,
F_{J-GS,\,\alpha\nu}^{\dag}\vert 0\rangle$ and $\prod_{\alpha\nu = c0,\,s1}\,
F_{-GS,\,\alpha\nu}^{\dag}\vert 0\rangle$, respectively. Furthermore, the ground-state
distribution ${\cal{N}}_{\alpha\nu}^{0}({\bar{q}}_j)$ is both such that ${\bar{q}}_j=q_j$
and ${\cal{N}}_{\alpha\nu}^{0}({\bar{q}}_j) =N_{\alpha\nu}^{0} (q_j)$, where
$N_{\alpha\nu}^{0} (q_j)$ is the ground-state bare-momentum distribution function given
in Eqs. (C.1)-(C.3) of Ref. \cite{I}. Thus, $\Delta
{\cal{N}}_{\alpha\nu}^{F}({\bar{q}}_j)=\Delta N_{\alpha\nu}^{F}(q_j)$ describes
$\alpha\nu =c0,\,s1$ pseudofermion addition to or removal from the {\it Fermi points}.
Moreover, $\Delta {\cal{N}}_{\alpha\nu}^{NF}({\bar{q}}_j) = \Delta
N_{\alpha\nu}^{NF}({\bar{q}}_j)$ describes $\alpha\nu$ pseudofermion creation and/or
annihilation away from the {\it Fermi points} for the $\alpha\nu =c0,\,s1$ branches and
creation of $\alpha\nu$ pseudofermions at canonical-momentum values such that
$\vert{\bar{q}}_j\vert<q^0_{\alpha\nu}$ for the $\alpha\nu\neq c0,\,s1$ branches, whereas
$\Delta {\cal{N}}_{\alpha\nu}^{F}({\bar{q}}_j) = \Delta N_{\alpha\nu}^{F}({\bar{q}}_j)$
describes creation of $\alpha\nu\neq c0,\,s1$ pseudofermions at the limiting
canonical-momentum values ${\bar{q}}_j=\pm q^0_{\alpha\nu}$. Finally, the deviation
$\Delta {\cal{N}}_{\alpha\nu}^{phF}({\bar{q}}_j)=\Delta N_{\alpha\nu}^{phF}(q_j)$
corresponds to low-energy and small-momentum $\alpha\nu= c0,\,s1$ pseudofermion
particle-hole processes. For $n=1$ (and/or $m=0$) the excitation subspace is such that
$\Delta {\cal{N}}_{c0}^{phF}({\bar{q}}_j)=\Delta N_{c0}^{phF}(q_j)=0$ (and/or $\Delta
{\cal{N}}_{s1}^{phF}({\bar{q}}_j)=\Delta N_{s1}^{phF}(q_j)=0$). The above deviations are
such that,
\begin{eqnarray}
\sum_{{{\bar{q}}_j}=-q_{\alpha\nu}^0}^{+q_{\alpha\nu}^0} \Delta
{\cal{N}}_{\alpha\nu}^{phF}({\bar{q}}_j) = 0 \, ; \hspace{0.5cm} \sum_{{{\bar{q}}_j}=
-q_{\alpha\nu}^0}^{+q_{\alpha\nu}^0} \Delta {\cal{N}}_{\alpha\nu}^{NF}({\bar{q}}_j) =
\Delta N^{NF}_{\alpha\nu} \, ; \hspace{0.5cm} \sum_{{{\bar{q}}_j}=
-q_{\alpha\nu}^0}^{+q_{\alpha\nu}^0} \Delta {\cal{N}}_{\alpha\nu}^{F}({\bar{q}}_j) =
\Delta N^{F}_{\alpha\nu}\, ; \hspace{0.5cm} \alpha\nu = c0,\,s1 \, . \nonumber
\label{DNc0s1}
\end{eqnarray}
Since for the $\alpha\nu\neq c0,\,s1$ pseudofermion branches there is no occupancy in the
initial ground state, the canonical-momentum distribution function is such that,
$\sum_{{{\bar{q}}_j}= -q_{\alpha\nu}^0}^{+q_{\alpha\nu}^0} \Delta
{\cal{N}}_{\alpha\nu}({\bar{q}}_j) = N_{\alpha\nu}^{NF}+N_{\alpha\nu}^{F}$. [See Eq.
(\ref{numbers}).]

The generator $F^{\dag}_{f,\,\alpha\nu}$ given in Eq. (\ref{excSD}) can be written as,
\begin{eqnarray}
F^{\dag}_{f,\,\alpha\nu} & = & F_{p-h,\,\alpha\nu}^{\dag}\,
F_{J-NF,\,\alpha\nu}^{\dag}\,F_{J-GS,\,\alpha\nu}^{\dag} \, ;
\hspace{0.5cm} \alpha\nu = c0,\,s1 \, , \nonumber \\
& = & F^{\dag}_{NF,\,\alpha\nu}\,F^{\dag}_{F,\,\alpha\nu} \, ; \hspace{0.5cm}
\alpha\nu\neq c0,\,s1 \, . \nonumber
\end{eqnarray}
The expressions of the generators $F_{J-NF,\,\alpha\nu}^{\dag}$,
$F^{\dag}_{p-h,\,\alpha\nu}$, and $F^{\dag}_{NF,\,\alpha\nu}$ in terms of pseudofermion
creation and annihilation operators are given in Eqs. (B.1), (B.4), and (B.5) of Ref.
\cite{V}, respectively, and that of the generator $F_{J-GS,\,\alpha\nu}^{\dag}$ is
provided in Eq. (\ref{GS-SD}), whereas the generator $F^{\dag}_{F,\,\alpha\nu}$ creates
$\alpha\nu\neq c0,\,s1$ pseudofermions at $q=\pm q^0_{\alpha\nu}$. The generators
$F^{\dag}_{p-h,\,\alpha\nu}\,F_{J-GS,\,\alpha\nu}^{\dag}$ and
$F_{J-NF,\,\alpha\nu}^{\dag}$ are associated with the deviations
${\cal{N}}_{\alpha\nu}^{ph}({\bar{q}}_j)$ and $\Delta
{\cal{N}}_{\alpha\nu}^{NF}({\bar{q}}_j)$ of Eq. (\ref{NanphTh}), respectively,
$F_{p-h,\,\alpha\nu}^{\dag}$ corresponds to the deviation $\Delta
{\cal{N}}_{\alpha\nu}^{phF}({\bar{q}}_j)$ of the same equation and thus generates the
low-energy and small-momentum $\alpha\nu =c0,\,s1$ pseudofermion particle-hole processes
(C). The operators $F^{\dag}_{NF,\,\alpha\nu}$ and $F^{\dag}_{F,\,\alpha\nu}$ refer to
the $\alpha\nu\neq c0,\,s1$ branches and are associated with the deviations $\Delta
{\cal{N}}_{\alpha\nu}^{NF}({\bar{q}}_j)$ and $\Delta
{\cal{N}}_{\alpha\nu}^{F}({\bar{q}}_j)$, respectively.

The precise expression of the spectral functions of Eq. (\ref{ABONj+-cp}) depends on the
specific form of the local operator ${\tilde{\Theta}}_{{{\cal{N}}_i},\,k}^{l}$, whose
expression includes contributions from all $\alpha\nu$ branches with finite pseudofermion
occupancy in the corresponding J-CPHS ensemble subspace. For each such a subspace that
operator expression and the corresponding coefficient $C_J$ appearing in Eqs.
(\ref{cjpm0})-(\ref{cjpm0-2}) have the following general form,
\begin{eqnarray}
{\tilde{\Theta}}_{{{\cal{N}}_i},\,j}^{l} & = & {1\over G_C }[\prod_{\alpha\nu}
\,{\tilde{\Theta}}_{j',\,\alpha\nu}^{l,i}] \, ; \hspace{0.5cm} C_J = e^{i\,j\Delta
P_J}\,[G_C/G_J] \, ; \hspace{0.5cm}  G_J = \prod_{\alpha\nu}\Bigl[\theta
({\cal{N}}_{\alpha\nu})(n^*_{\alpha\nu})^{{\cal{N}}_{\alpha\nu}-1\over 2} + [1 - \theta
({\cal{N}}_{\alpha\nu})]\Bigr] \, , \nonumber \\
{\cal{N}}_{\alpha\nu} & = & \vert\Delta N_{\alpha\nu}^{NF}\vert + 2N^{phNF}_{\alpha\nu}
\, , \hspace{0.25cm} \alpha\nu = c0,\,s1 \, ; \hspace{0.50cm} {\cal{N}}_{\alpha\nu}=
N^{NF}_{\alpha\nu} \, , \hspace{0.25cm} \alpha\nu\neq c0,\,s1 \, , \label{Than}
\end{eqnarray}
where the values of the positive real coefficient $G_C$ and momentum deviation $\Delta
P_J$ are given below and we recall that $j' = j n^*_{\alpha\nu}$ denotes the integer
number closest to $j n^*_{\alpha\nu}$. In equation (\ref{Than}) and in the remaining of
this paper $\theta (x)$ is such that $\theta (x)=1$ for $x> 0$ and $\theta (x)=0$ for
$x\leq 0$.

The value of the coefficient $G_C$ appearing in the general expression for the operator
${\tilde{\Theta}}_{{{\cal{N}}_i},\,j}^{l}$ given in Eq. (\ref{Than}) is the same for all
J-CPHS ensemble subspaces belonging to a given CPHS ensemble subspace, whereas that of
$G_J$ and thus of $\vert C_J\vert$ is specific to each J-CPHS ensemble subspace.
Furthermore, when the expression of the local operator ${\hat{O}}_{{\cal{N}},\,j}^{l}$ of
Eq. (\ref{ABONj}) is independent of $U/t$, the same occurs for the related operators
${\hat{\Theta}}_{{\cal{N}},\,j}^{l}$ and ${\tilde{\Theta}}_{{\cal{N}}_0,\,j}^l$ of Eqs.
(\ref{Th})-(\ref{OL-0}) and (\ref{ONjtil}), respectively. It follows that in the case of
the $i=0$ operator ${\tilde{\Theta}}_{{\cal{N}}_0,\,j}^l$, the value of the coefficient
$G_C$ appearing in its expression given in Eq. (\ref{Than}) is also independent of $U/t$.
Fortunately, for the dominant CPHS ensemble subspaces considered in the previous section,
such a value can be found from analysis of the problem for $U/t=0$ or $U/t=\infty$. One
then finds that $G_C=1$ in the ${\tilde{\Theta}}_{{{\cal{N}}_0},\,j}^{l}={1\over G_C
}[\prod_{\alpha\nu} \,{\tilde{\Theta}}_{j',\,\alpha\nu}^{l,0}]$ expression corresponding
to the dominant CPHS ensemble subspaces.

The operator ${\tilde{\Theta}}_{j',\,\alpha\nu}^{l,i}$ appearing in Eq. (\ref{Than}) has
the following general form for the $\alpha\nu$ branches with finite pseudofermion
occupancy in the J-CPHS subspace,
\begin{eqnarray}
{\tilde{\Theta}}_{j',\,\alpha\nu}^{l,i} & = & e^{-i\Delta
P^0_{\alpha\nu}j'a^0_{\alpha\nu}}\,{\tilde{\Theta}}_{j',\,\alpha\nu}^{l,NF,i}\,
{\tilde{\Theta}}_{\alpha\nu}^{l,F,i} \, . \nonumber
\end{eqnarray}
Here the operators ${\tilde{\Theta}}_{j',\,\alpha\nu}^{l,NF,i}$ and
${\tilde{\Theta}}_{\alpha\nu}^{l,F,i}$ are associated with the elementary processes (A)
and (B), respectively, and the {\it phase-factor momentum} $l\Delta P^0_{\alpha\nu}$ is
given below. Considering that $[x_{j'}-x_j]=0$, it is such that $-i\sum_{\alpha\nu}\Delta
P^0_{\alpha\nu}j'a^0_{\alpha\nu}=-ij\Delta P_J$. Here the summation $\sum_{\alpha\nu}$ is
over the $\alpha\nu$ branches with pseudofermion occupancy in the J-CPHS ensemble
subspace and $\Delta P_J$ is the momentum deviation of Eq. (\ref{Than}), which reads,
\begin{eqnarray}
\Delta P_J & = & \sum_{\alpha\nu}\Delta P^0_{\alpha\nu} \, . \nonumber
\end{eqnarray}

Let us proceed by studying the phase factor $e^{-i\Delta
P^0_{\alpha\nu}j'a^0_{\alpha\nu}}$ and operators ${\tilde{\Theta}}_{\alpha\nu}^{l,F,i}$
and ${\tilde{\Theta}}_{j',\,\alpha\nu}^{l,NF,i}$ whose product gives the operator
${\tilde{\Theta}}_{j',\,\alpha\nu}^{l,i}$. We start by characterizing for each
$\alpha\nu$ branch the elementary processes that originate the phase factor $e^{-i\Delta
P^0_{\alpha\nu}j'a^0_{\alpha\nu}}$. The expression of the local operator
${\tilde{\Theta}}_{j',\,\alpha\nu}^{l,i}$ does not involve the generator of the
elementary processes (C), $F_{p-h,\,\alpha\nu}^{\dag}$, but provides the momentum
$l\Delta P^{phF}_{\alpha\nu}$ for such processes through a phase factor, $e^{-i\Delta
P^{phF}_{\alpha\nu}j'a^0_{\alpha\nu}}$. Here $l\Delta
P^{phF}_{\alpha\nu}=l[2\pi/L]\,[m_{\alpha\nu,\,+1}-m_{\alpha\nu,\,-1}]$ is the small
momentum deviation generated by the $\alpha\nu =c0,\,s1$ pseudofermion particle-hole
elementary processes (C) and $m_{\alpha\nu,\,\pm 1}$ is the number of such processes of
momentum $\pm[2\pi/L]$ in the vicinity of the Fermi points $\pm q^0_{F\alpha\nu}$.
Furthermore, each $\alpha\nu=c0,\,s1$ pseudofermion created or annihilated at the {\it
Fermi point} $\iota\,q^0_{F\alpha\nu}$ by the elementary processes (B) contributes with a
phase factor $e^{-i\iota\,q^0_{F\alpha\nu}j'a^0_{\alpha\nu}}$ or
$e^{+i\iota\,q^0_{F\alpha\nu}j'a^0_{\alpha\nu}}$, respectively, where $\iota =\pm 1$.
This leads to a phase factor $e^{-i2q^0_{F\alpha\nu}\Delta
J^{0,F}_{\alpha\nu}j'a^0_{\alpha\nu}}$, where $2\Delta J^{0,F}_{\alpha\nu}=\Delta
N^{0,F}_{\alpha\nu ,\,+1}-\Delta N^{0,F}_{\alpha\nu ,\,-1}$ and the number deviation
$\Delta N^{0,F}_{\alpha\nu ,\,\pm 1}$ is that defined in Eq. (\ref{numbers}). Moreover,
each $c\nu\neq c0$ and $s\nu\neq s1$ FP scattering center created by the elementary
processes (B) contributes with a phase factor $e^{-i\iota\,q^0_{Fc0}j'a^0_{c0}}$ and
$e^{-i\iota\,q^0_{Fc0}j'a^0_{c0}}\,e^{+i\iota\,2q^0_{Fs1}j'a^0_{s1}}$, respectively. On
the other hand, the scattering-less bare-momentum shift contributes with a phase factor
$e^{-i[Q^0_{\alpha\nu}/L]j'a^0_{c\nu}}$ for each of the $N^0_{\alpha\nu}$ pseudofermions
of the initial ground state, what gives $[e^{-i[Q^0_{\alpha\nu}/L]
j'a^0_{c\nu}}]^{N^0_{\alpha\nu}}=
e^{-iq^0_{F\alpha\nu}[Q^0_{\alpha\nu}/\pi]j'a^0_{c\nu}}$ with $\alpha\nu=c0,\,s1$. Adding
all these contributions leads to the above net phase factor $e^{-i\Delta
P^0_{\alpha\nu}j'a^0_{\alpha\nu}}$ for the $\alpha\nu=c0,\,s1$ branches whose
phase-factor momentum reads $l\Delta P^0_{\alpha\nu}=l[\Delta P^{phF}_{\alpha\nu}+\Delta
P^F_{\alpha\nu}]$. For densities in the ranges $0<n<1$ and $0<m<n$, the momentum $\Delta
P^F_{\alpha\nu}$ appearing in that phase factor is given by $\Delta
P^F_{c0}=4k_F[\,\Delta J^F_{c0} + \sum_{\nu =1}^{\infty}\,J^F_{c\nu} + \sum_{\nu
=2}^{\infty}\,J^F_{s\nu}]$ and $\Delta P^F_{s1}=2k_{F\downarrow}[\,\Delta J^F_{s1} -
2\sum_{\nu =2}^{\infty}\,J^F_{s\nu}]$ for the $\alpha\nu = c0$ and $\alpha\nu =s1$
branches, respectively. It results from the current contributions associated with the
$c0$ and $s1$ {\it Fermi points} $\pm 2k_F$ and $\pm k_{F\downarrow}$, respectively.
Moreover, each $c\nu\neq c0$ pseudofermion created by the elementary processes (A)
contributes with a phase factor $e^{-i(1+\nu)\pi j'a^0_{c\nu}}$ what leads to a net phase
factor $e^{-i\Delta P^0_{c\nu}j'a^0_{c\nu}}$ for the $c\nu\neq c0$ branches such that
$\Delta P^0_{c\nu}= (1+\nu)\pi {\cal{N}}_{c\nu}$. Finally, the phase-factor momentum
vanishes for the $s\nu\neq s1$ branches. Thus, the phase-factor momenta contributing to
$\Delta P_J$ read,
\begin{eqnarray}
l\Delta P^0_{\alpha\nu} = l[\Delta P^{phF}_{\alpha\nu}+\Delta P^F_{\alpha\nu}] \, ,
\hspace{0.25cm} \alpha\nu = c0,\,s1 \, ; \hspace{0.50cm} l\Delta P^0_{c\nu} = l(1+\nu)\pi
{\cal{N}}_{c\nu} \, , \hspace{0.25cm} c\nu\neq c0 \, ; \hspace{0.50cm} l\Delta P^0_{s\nu}
= 0 \, , \hspace{0.25cm} s\nu\neq s1 \, . \nonumber
\end{eqnarray}

Next we consider the operator ${\tilde{\Theta}}_{\alpha\nu}^{l,F,i}$. For the $c\nu\neq
c0$ and $s\nu\neq s1$ branches, that operator creates $2\nu N^F_{c\nu}$ independent
$-1/2$ holons of momentum $\pi$ and $2\nu N^F_{s\nu}$ independent $-1/2$ spinons of
momentum zero, respectively. (The only effect of the creation of the corresponding
$N^F_{c\nu}$ $c\nu\neq c0$ FP scattering centers and $N^F_{s\nu}$ $s\nu\neq s1$ FP
scattering centers \cite{V}, respectively, is the above contribution to the phase factor
$e^{-i\Delta P^0_{c0}j'a^0_{c0}}$ and $e^{-i\Delta P^0_{c0}j'a^0_{c0}}e^{-i\Delta
P^0_{s1}j'a^0_{s1}}$, respectively.) For the $\alpha\nu = c0,\,s1$ branches the operator
${\tilde{\Theta}}_{\alpha\nu}^{l,F,i}$ is such that,
\begin{eqnarray}
e^{-i2q^0_{\alpha\nu}\Delta J^{0,F}_{\alpha\nu}j'a^0_{\alpha\nu}}\,
{\tilde{\Theta}}_{\alpha\nu}^{l,F,i}\, & = & \left({1\over
N^*_{\alpha\nu}}\right)^{\vert\Delta N^F_{\alpha\nu}\vert\over 2}\prod_{\iota =\pm
1}\Bigl\{\,\Theta (\Delta N^F_{\alpha\nu,\,\iota})\prod_{i'=1}^{\Delta
N^F_{\alpha\nu,\,\iota}}\Bigl[\sum_{j''=1}^{N^*_{\alpha\nu}}\, e^{-i\iota
(q^0_{\alpha\nu} +2\pi\,i'/L)a_{\alpha\nu}^0(j'-j'')}
f^{\dag}_{x_{j''},\,\alpha\nu}\Bigr] \nonumber \\
& + & \Theta (-\Delta N^F_{\alpha\nu,\,\iota})\prod_{i'=0}^{\vert\Delta
N^F_{\alpha\nu,\,\iota}\vert-1}\Bigl[\sum_{j''=1}^{N^*_{\alpha\nu}}\, e^{i\iota
(q^0_{\alpha\nu} -2\pi\,i'/L)a_{\alpha\nu}^0(j'-j'')}
f_{x_{j''},\,\alpha\nu}\Bigr]\Bigr\} \, ; \hspace{0.25cm} \alpha\nu = c0,\,s1 \, .
\nonumber
\end{eqnarray}
Thus, that operator can be written as,
\begin{eqnarray}
{\tilde{\Theta}}_{\alpha\nu}^{l,F,i}\, & = & \prod_{\iota =\pm 1}\Bigl\{\,\Theta (\Delta
N^F_{\alpha\nu,\,\iota})\prod_{i'=1}^{\Delta N^F_{\alpha\nu,\,\iota}} f^{\dag}_{\iota
(q^0_{\alpha\nu} +2\pi\,i'/L),\,\alpha\nu} + \Theta (-\Delta
N^F_{\alpha\nu,\,\iota})\prod_{i'=0}^{\vert\Delta N^F_{\alpha\nu,\,\iota}\vert-1}
f_{\iota (q^0_{\alpha\nu} -2\pi\,i'/L),\,\alpha\nu}\Bigr\} \, ; \hspace{0.25cm} \alpha\nu
= c0,\,s1 \, , \nonumber
\end{eqnarray}
where we omitted corrections of order $1/L$ to the momentum value $q^0_{\alpha\nu}$
appearing in the phase factor. However, these corrections must be considered in the
momentum of the pseudofermion operators. In the above two equations the pseudofermion
operators correspond to spatial and canonical-momentum variables, respectively, and in
the argument of the exponentials appearing in these equations and in other equations
given below $i$ is the usual imaginary number. (It is not the index $i$ of the operator
${\tilde{\Theta}}_{\alpha\nu}^{l,F,i}$, whereas $i'$ is a summation index.) Moreover, in
the above equations and in the remaining of this paper $\Theta (x)$ is such that $\Theta
(x)=1$ for $x\geq 0$ and $\Theta (x)=0$ for $x< 0$.

We finish the study of the operator ${\tilde{\Theta}}_{j',\,\alpha\nu}^{l,i}$ by
considering the operator ${\tilde{\Theta}}_{j',\,\alpha\nu}^{l,NF,i}$ associated with the
elementary processes (A), whose expression refers to a given J-CPHS subspace. In order to
arrive to that expression in terms of the local $\alpha\nu$ pseudofermion creation and
annihilation operators, we recall that our study refers to spectral functions of form
(\ref{ABONj}) whose operator ${\hat{O}}_{{\cal{N}},\,j}^{l}$ expression involves
${\cal{N}}$ elementary electronic operators which create or annihilate electrons in a
compact domain of lattice sites. For such an operator the general expression of the
corresponding operators ${\tilde{\Theta}}_{j',\,\alpha\nu}^{l,NF,i}$ involves
${\cal{N}}_{\alpha\nu}$ local $\alpha\nu$ pseudofermion creation and annihilation
operators for the $\alpha\nu= c0,\,s1$ branches (and creation operators for the
$\alpha\nu\neq c0,\,s1$ branches) which refer to a compact domain of
${\cal{N}}_{\alpha\nu}$ effective $\alpha\nu$ lattice sites. The operators
(\ref{cjpm0})-(\ref{cjpm0-2}) involve the product of operators whose expressions involve
elementary operators of a single $\alpha\nu$ branch and are particular examples of such a
general expression, which has the following form,
\begin{eqnarray}
{\tilde{\Theta}}_{j',\,\alpha\nu}^{l,NF,i} & = &
(n^*_{\alpha\nu})^{{\cal{N}}_{\alpha\nu}-1\over 2} \,\Bigl[\Theta (\Delta
N_{\alpha\nu}^{NF})\prod_{j''=j'}^{\Delta
N_{\alpha\nu}^{NF}+N^{phNF}_{\alpha\nu}+j'-1}\prod_{j'''=\Delta
N_{\alpha\nu}^{NF}+N^{phNF}_{\alpha\nu}+j'}^{{\cal{N}}_{\alpha\nu}+j'-1}
f^{\dag}_{x_{j''},\,\alpha\nu}\,f_{x_{j'''},\,\alpha\nu} \nonumber \\
& + & \theta (-\Delta N_{\alpha\nu}^{NF})\prod_{j''=j'}^{\vert\Delta
N_{\alpha\nu}^{NF}\vert+N^{phNF}_{\alpha\nu}+j'-1}\prod_{j'''=\vert\Delta
N_{\alpha\nu}^{NF}\vert+N^{phNF}_{\alpha\nu}+j'}^{{\cal{N}}_{\alpha\nu}+j'-1}
f_{x_{j''},\,\alpha\nu}\,f^{\dag}_{x_{j'''},\,\alpha\nu}\Bigr] \, ; \hspace{0.5cm}
\alpha\nu = c0,\,s1 \, , \nonumber
\end{eqnarray}
and
\begin{eqnarray}
{\tilde{\Theta}}_{j',\,\alpha\nu}^{l,NF,i} & = &
(n^*_{\alpha\nu})^{{\cal{N}}_{\alpha\nu}-1\over 2}
\,\prod_{j''=j'}^{{\cal{N}}_{\alpha\nu}+j'-1} f^{\dag}_{x_{j''},\,\alpha\nu} \, ;
\hspace{0.25cm} c\nu\neq c0 \, , \hspace{0.15cm} n<1 \, ; \hspace{0.25cm} s\nu\neq s1 \,
, \hspace{0.15cm} m<n \, . \nonumber
\end{eqnarray}
Here the spatial coordinates $x_0$, $x_1$,...,$x_{{\cal{N}}_{\alpha\nu}-1}$ correspond to
the compact domain of ${\cal{N}}_{\alpha\nu}$ effective $\alpha\nu$ lattice sites where
the number ${\cal{N}}_{\alpha\nu}$ is given in Eq. (\ref{Than}).

Let us consider the operator ${\tilde{\Theta}}_{k,\,\alpha\nu}^{l,i}$, which is the
Fourier transform of the above local operator ${\tilde{\Theta}}_{j',\,\alpha\nu}^{l,i}$,
\begin{eqnarray}
{\tilde{\Theta}}_{k,\,\alpha\nu}^{l,i} =
{1\over{\sqrt{N^*_{\alpha\nu}}}}\sum_{j'=1}^{N^*_{\alpha\nu}}\,e^{+i\,lkj'a^0_{\alpha\nu}}\,
{\tilde{\Theta}}_{j',\,\alpha\nu}^{l,i} \, . \nonumber
\end{eqnarray}
This operator can be expressed in terms of the following momentum convolution,
\begin{eqnarray}
{\tilde{\Theta}}_{k,\,\alpha\nu}^{l,i} & = &
\sum_{k'}{\tilde{\Theta}}_{k-k',\,\alpha\nu}^{l,NF,i}\,{\tilde{\Theta}}_{\alpha\nu}^{l,F,i}
\,\delta_{k',\,l\Delta P^0_{\alpha\nu}}= {\tilde{\Theta}}_{k-l\Delta
P^0_{\alpha\nu},\,\alpha\nu}^{l,NF,i}\,{\tilde{\Theta}}_{\alpha\nu}^{l,F,i} \, ,
\nonumber
\end{eqnarray}
where the operator ${\tilde{\Theta}}_{\alpha\nu}^{l,F,i}$ was given above and
${\tilde{\Theta}}_{k,\,\alpha\nu}^{l,NF,i}$ is the Fourier transform of the local
operator ${\tilde{\Theta}}_{j',\,\alpha\nu}^{l,NF,i}$. It reads,
\begin{eqnarray}
{\tilde{\Theta}}_{k,\,\alpha\nu}^{l,NF,i} & = & \left({1\over
N_a}\right)^{{\cal{N}}_{\alpha\nu}-1\over 2} \Bigl\{\,\Theta (\Delta
N_{\alpha\nu}^{NF})\Bigl[\prod_{i'=0}^{\Delta
N_{\alpha\nu}^{NF}+N^{phNF}_{\alpha\nu}-1}\prod_{i''=\Delta
N_{\alpha\nu}^{NF}+N^{phNF}_{\alpha\nu}}^{{\cal{N}}_{\alpha\nu}-1}
\Bigl(\sum_{{\bar{q}}_{i'}=-q^0_{\alpha\nu}}^{+q^0_{\alpha\nu}}\,
e^{-ii'a_{\alpha\nu}^0{\bar{q}}_{i'}}\,f^{\dag
}_{{\bar{q}}_{i'},\,\alpha\nu}\Bigr)\nonumber \\
& \times & \Bigl(\sum_{{\bar{q}}_{i''}=-q^0_{\alpha\nu}}^{+q^0_{\alpha\nu}}\,
e^{ii''a_{\alpha\nu}^0{\bar{q}}_{i''}}\,f_{{\bar{q}}_{i''},\,\alpha\nu}\Bigr)\Bigr]
\delta_{k,\,l[\sum_{i'=0}^{\Delta
N_{\alpha\nu}^{NF}+N^{phNF}_{\alpha\nu}-1}{\bar{q}}_{i'}-\sum_{i''=\Delta
N_{\alpha\nu}^{NF}+N^{phNF}_{\alpha\nu}}^{{\cal{N}}_{\alpha\nu}-1}{\bar{q}}_{i''}]}
\nonumber \\
& + & \theta (-\Delta N_{\alpha\nu}^{NF})\Bigl[\prod_{i'=0}^{\vert\Delta
N_{\alpha\nu}^{NF}\vert+N^{phNF}_{\alpha\nu}-1}\prod_{i''=\vert\Delta
N_{\alpha\nu}^{NF}\vert+N^{phNF}_{\alpha\nu}}^{{\cal{N}}_{\alpha\nu}-1}
\Bigl(\sum_{{\bar{q}}_{i'}=-q^0_{\alpha\nu}}^{+q^0_{\alpha\nu}}\,
e^{ii'a_{\alpha\nu}^0{\bar{q}}_{i'}}\,f_{{\bar{q}}_{i'},\,\alpha\nu}\Bigr)\nonumber
\\
& \times & \Bigl(\sum_{{\bar{q}}_{i''}=-q^0_{\alpha\nu}}^{+q^0_{\alpha\nu}}\,
e^{-ii''a_{\alpha\nu}^0{\bar{q}}_{i''}}\,f^{\dag}_{{\bar{q}}_{i''},\,\alpha\nu}\Bigr)\Bigr]\nonumber
\\
& \times & \delta_{k,\,l[-\sum_{i'=0}^{\vert\Delta
N_{\alpha\nu}^{NF}\vert+N^{phNF}_{\alpha\nu}-1}{\bar{q}}_{i'}+\sum_{i''=\vert\Delta
N_{\alpha\nu}^{NF}\vert+N^{phNF}_{\alpha\nu}}^{{\cal{N}}_{\alpha\nu}-1}{\bar{q}}_{i''}]}
\Bigr\} \, ; \hspace{0.5cm} \alpha\nu = c0,\,s1 \, , \nonumber \\
{\tilde{\Theta}}_{k,\,\alpha\nu}^{l,NF,i} & = & \left({1\over
N_a}\right)^{{\cal{N}}_{\alpha\nu}-1\over 2}\Bigl[\prod_{i'=0}^{{\cal{N}}_{\alpha\nu}-1}
\sum_{{\bar{q}}_{i'}=-q^0_{\alpha\nu}}^{+q^0_{\alpha\nu}}\,
e^{-ii'a_{\alpha\nu}^0{\bar{q}}_{i'}}\,
f^{\dag}_{{\bar{q}}_{i'},\,\alpha\nu}\Bigr] \nonumber \\
& \times &
\delta_{k,\,l[c_{\alpha\nu}\sum_{i'=0}^{{\cal{N}}_{\alpha\nu}-1}{\bar{q}}_{i'}]}\, ;
\hspace{0.5cm} c\nu\neq c0 \, , \hspace{0.25cm} n<1 \, ; \hspace{0.5cm} s\nu\neq s1 \, ,
\hspace{0.25cm} m>0 \, . \label{Thetas}
\end{eqnarray}
Here the set ${\bar{q}}_{0}$, ${\bar{q}}_{1}$,
${\bar{q}}_{2}$,...,${\bar{q}}_{{\cal{N}}_{\alpha\nu}-1}$ refers to
${\cal{N}}_{\alpha\nu}$ summation canonical-momentum variables associated with the
$\alpha\nu$ pseudofermion bands. We note that the canonical-momentum values in the
Kr\"onecker $\delta$'s of Eq. (\ref{Thetas}) run under the summations but not under the
products appearing in that equation. For the $\alpha\nu\neq c0,\,s1$ branches, the
${\cal{N}}_{\alpha\nu}$-$\alpha\nu$ pseudofermion operator
${\tilde{\Theta}}_{k,\,\alpha\nu}^{l,NF,i}$ creates and annihilates $\vert\Delta
N_{\alpha\nu}^{NF}\vert$ $\alpha\nu$ pseudofermions when $\Delta N_{\alpha\nu}^{NF} >0$
and $\Delta N_{\alpha\nu}^{NF} <0$, respectively, and generates
$N^{phNF}_{\alpha\nu}=0,1,...$ finite-momentum and finite-energy $\alpha\nu$
pseudofermion particle-hole processes. For the $\alpha\nu\neq c0,\,s1$ branches, it
creates $N_{\alpha\nu}^{NF}$ $\alpha\nu$ pseudofermions of bare-momentum $q$ such that
$\vert q\vert<q^0_{\alpha\nu}$.

In the case of the $\alpha\nu\neq c0,\,s1$ branches, the general expression given in Eq.
(\ref{Thetas}) for the operator ${\tilde{\Theta}}_{k,\,\alpha\nu}^{l,NF,i}$ is valid for
densities such that the corresponding ratios $n^*_{\alpha\nu}=N^*_{\alpha\nu}/N_a$ have
finite values. For the $c\nu\neq c0$ (and $s\nu\neq s1$) pseudofermion branches and
electronic density $n=1$ (and spin density $m=0$) all pseudofermions separate into $2\nu$
independent $-1/2$ holons (and $2\nu$ independent $-1/2$ spinons). Therefore, the
operator given in Eq. (\ref{Thetas}) does not exist. Moreover, for $n=1$ (and $m=0$) the
above generator $F_{p-h,\,c0}^{\dag}$ (and $F_{p-h,\,s1}^{\dag}$) of the elementary
processes (C) reduces to $F_{p-h,\,c0}^{\dag}=1$ (and $F_{p-h,\,s1}^{\dag}=1$). However,
our theory also applies for electronic density $n=1$ (and spin density $m=0$), provided
that we consider the corresponding restrictions in the $c0$ (and $s1$) excitation
spectrum and take into account that
${\tilde{\Theta}}_{k,\,\alpha\nu}^{l,i}={\tilde{\Theta}}_{k,\,\alpha\nu}^{l,NF,i}$, for
the $c\nu\neq c0$ (and $s\nu\neq s1$) branches.

For the $\alpha\nu =c0,\, s1$ branches the pseudofermion weight distribution involves the
following matrix element,
\begin{eqnarray}
& & \langle 0\vert F_{J-GS,\,\alpha\nu}\,F_{J-NF,\,\alpha\nu}\,F_{p-h,\,\alpha\nu}\,
{\tilde{\Theta}}_{k,\,\alpha\nu}^{l,i}\,F_{GS,\,\alpha\nu}^{\dag}\vert 0\rangle \, ;
\hspace{0.5cm} \alpha\nu = c0,\,s1  \, . \nonumber
\end{eqnarray}
Our study refers to very large values of $L$ when the commutator
$[F^{\dag}_{p-h,\,\alpha\nu},\,F_{J-NF,\,\alpha\nu}^{\dag}]=0$ vanishes and the operator
${\tilde{\Theta}}_{\alpha\nu}^{l,F,i}$ is such that
${\tilde{\Theta}}_{\alpha\nu}^{l,F,i}\,F_{GS,\,\alpha\nu}^{\dag}=F_{-GS,\,\alpha\nu}^{\dag}$
and thus the matrix element can be rewritten as,
\begin{eqnarray}
& & \langle 0\vert F_{J-GS,\,\alpha\nu}\,F_{p-h,\,\alpha\nu}\,F_{J-NF,\,\alpha\nu}\,
{\tilde{\Theta}}_{k-l\Delta
P^0_{\alpha\nu},\,\alpha\nu}^{l,NF,i}\,F_{-GS,\,\alpha\nu}^{\dag}\vert 0\rangle \, ;
\hspace{0.5cm} \alpha\nu = c0,\,s1  \, . \nonumber
\end{eqnarray}

When applying the generators $F_{f,\,L,\,\alpha\nu}^{\dag}$ and
$F_{GS,\,\alpha\nu}^{\dag}$ of Eqs. (\ref{excSD}) and (\ref{GS-SD}), respectively, onto
the pseudofermion vacuum to construct a given energy eigenstate, the set of $\alpha\nu$
band discrete canonical-momentum values $\{{\bar{q}}_j\}$ of the pseudofermion operators
$f^{\dag }_{{\bar{q}}_j,\,\alpha\nu}$ in the expressions of these generators are those of
the CPHS ensemble subspace which that state belongs to. This rule applies when one
considers the generators of the full energy eigenstates. (Below we express each J-CPHS as
a direct product of subspaces. To reach the correct final results, such a rule does not
apply to some of the states which span such direct-product subspaces.) The same occurs
with the discrete canonical-momentum values of the pseudofermion creation and
annihilation operators of the expression of any operator ${\tilde{\Theta}}$ when it acts
onto a given energy eigenstate. For instance, let $\vert\beta\rangle$ and
$\vert\beta'\rangle$ be energy eigenstates. Thus, the discrete canonical-momentum values
of the pseudofermion creation and annihilation operators of the expression of the
operators ${\tilde{\Theta}}^{\dag}$ and ${\tilde{\Theta}}$ in
${\tilde{\Theta}}^{\dag}\vert\beta'\rangle$ and ${\tilde{\Theta}}\vert\beta\rangle$ are
those of the CPHS ensemble subspace which the states $\vert\beta'\rangle$ and
$\vert\beta\rangle$ belong to, respectively. An important property for our theory is that
for $L$ large both choices lead to the same value for the matrix element
$\langle\beta\vert {\tilde{\Theta}}\vert\beta'\rangle$.

The operator ${\tilde{\Theta}}_{k,\,\alpha\nu}^{l,NF,i}$ of Eq. (\ref{Thetas}) is for the
$\alpha\nu = c0,\,s1$ branches such that the commutator
$[F_{p-h,\,\alpha\nu},\,{\tilde{\Theta}}_{k,\,\alpha\nu}^{l,NF,i}]=0$ vanishes when it
acts onto the CPHS ensemble subspace which the corresponding excited state belongs to.
Furthermore, there occurs a full overlap of that operator with the generator
$F_{J-NF,\,\alpha\nu}$ and for the $\alpha\nu\neq c0,\,s1$ branches there occurs a full
overlap of the operator ${\tilde{\Theta}}_{k,\,\alpha\nu}^{l,NF,i}$ given in the second
expression of Eq. (\ref{Thetas}) with the generator $F_{NF,\,\alpha\nu}$. The latter full
overlap results from the lack of $\alpha\nu\neq c0,\,s1$ pseudofermion occupancy of the
initial ground state.

A J-CPHS ensemble subspace can be expressed as the direct product of subspaces, one for
each $\alpha\nu$ branch pseudofermion occupancy. In the particular case of the $\alpha\nu
=c0,\,s1$ branches the low-energy and high-energy physics separate provided that $L$ is
large and one can define two of such product subspaces for each branch. They are
associated with the excitation occupancy configurations generated by the finite-energy
elementary processes (A) and low-energy elementary processes (B,C), respectively. We call
$p-h,\alpha\nu = c0,\,s1$ branch subspace the latter low-energy subspace. Thus, the
number of product subspaces equals the number of $\alpha\nu$ branches with finite
pseudofermion occupancy in the J-CPHS ensemble subspace plus two. Finally, for some
J-CPHS ensemble subspaces the direct product also includes the independent $-1/2$ holon
subspace and independent $-1/2$ spinon subspace. For such subspaces the generator
$F_{F,\,\alpha\nu}^{\dag}$ either creates $N^F_{\alpha\nu}$ $\alpha\nu$ pseudofermions
with limiting bare-momentum values $q=\pm q^0_{\alpha\nu}$ or reads
$F_{F,\,\alpha\nu}^{\dag}=1$ when $N^F_{\alpha\nu}=0$ and thus
$F_{f,\,\alpha\nu}^{\dag}=F_{NF,\,\alpha\nu}^{\dag}$ (or $F_{f,\,\alpha\nu}^{\dag}$ does
not exist if $n=1$ and $\alpha\nu =c\nu$ or $m=0$ and $\alpha\nu = s\nu$). The states
which span the $\alpha\nu$ branch direct-product subspaces and $p-h,\alpha\nu = c0,\,s1$
branch direct-product subspaces have the following form,
\begin{eqnarray}
\vert f.L;\,\alpha\nu\rangle & \equiv & F_{J-NF,\,\alpha\nu}^{\dag}\vert GS\rangle \, ;
\hspace{0.25cm} \alpha\nu = c0,\,s1 \, ; \hspace{0.5cm} \vert f.L;\,\alpha\nu\rangle
\equiv F_{NF,\,\alpha\nu}^{\dag}\vert GS\rangle \, ; \hspace{0.25cm} \alpha\nu \neq
c0,\,s1 \, ,
\nonumber \\
\vert f.L;\,p-h,\alpha\nu\rangle  & \equiv &
F_{p-h,\,\alpha\nu}^{\dag}\,F_{J-GS,\,\alpha\nu}^{\dag}\vert 0\rangle \, ;
\hspace{0.25cm} \alpha\nu= c0,\,s1 \, . \nonumber
\end{eqnarray}
Below we express the spectral functions as convolutions of $p-h,\,\alpha\nu$ and
$\alpha\nu$ pseudofermion spectral functions. However, in order to reach the same
spectral-weight distributions as by use of the above matrix
elements, it turns out that:\\ \\
(i) When applying the generators $F_{J-NF,\,\alpha\nu}^{\dag}$ and
$F_{NF,\,\alpha\nu}^{\dag}$ onto the ground state $\vert GS\rangle$ to construct a given
$\alpha\nu$ branch direct-product-subspace state, the set of the $\alpha\nu$ band
discrete canonical-momentum values $\{{\bar{q}}_j\}$ of the pseudofermion operators
$f^{\dag }_{{\bar{q}}_j,\,\alpha\nu}$ in the expressions of these
generators must be those of the ground-state CPHS ensemble subspace;\\ \\
(ii) When applying the generator
$F_{p-h,\,\alpha\nu}^{\dag}\,F_{J-GS,\,\alpha\nu}^{\dag}$ onto the pseudofermion vacuum
$\vert 0\rangle$ to construct a given $p-h,\alpha\nu = c0,\,s1$ branch
direct-product-subspace state, the set of the $\alpha\nu$ band discrete
canonical-momentum values $\{{\bar{q}}_j\}$ of the pseudofermion operators $f^{\dag
}_{{\bar{q}}_j,\,\alpha\nu}$ in the expressions of
that generator must be those of the corresponding excited energy eigenstate.\\

The property (i) ensures that the above full matrix-element overlaps are reproduced.
Furthermore, properties (i) and (ii) also ensure that the contribution from the
unconventional orthogonality catastrophe matrix-element overlap discussed below is not
counted twice.

Below we introduce the $\alpha\nu = c0,\,s1$ pseudofermion spectral functions and
$p-h,\alpha\nu = c0,\,s1$ pseudofermion spectral functions which involve the operators
${\tilde{\Theta}}_{k-l\Delta P^F_{\alpha\nu},\,\alpha\nu}^{l,NF,i}$ and
${\tilde{\Theta}}_{\alpha\nu}^{l,F,i} \,\delta_{k',\,l\Delta P^{phF}_{\alpha\nu}}$,
respectively. The momentum convolution of these two operators leads to the correct
expression for the operator ${\tilde{\Theta}}_{k,\,\alpha\nu}^{l,i}$ such that
${\tilde{\Theta}}_{k,\,\alpha\nu}^{l,i} = \sum_{k'}{\tilde{\Theta}}_{k-l\Delta
P^F_{\alpha\nu}-k',\,\alpha\nu}^{l,NF,i}\,{\tilde{\Theta}}_{\alpha\nu}^{l,F,i}
\,\delta_{k',\,l\Delta P^{phF}_{\alpha\nu}}= {\tilde{\Theta}}_{k-l\Delta
P^0_{\alpha\nu},\,\alpha\nu}^{l,NF,i}\,{\tilde{\Theta}}_{\alpha\nu}^{l,F,i}$. In turn,
the $\alpha\nu\neq c0,\,s1$ pseudofermion spectral functions considered below correspond
to the operators ${\tilde{\Theta}}_{k-l\Delta P^0_{\alpha\nu},\,\alpha\nu}^{l,NF,i}$,
which refer to the $\alpha\nu\neq c0,\,s1$ branch direct-product subspaces. For the
latter branches, the pseudofermion spectral function is associated with the
${\cal{N}}_{\alpha\nu}=N^{NF}_{\alpha\nu}>0$ $\alpha\nu$ pseudofermions created by the
processes (A), whereas the $N^{F}_{\alpha\nu}$ $\alpha\nu$ pseudofermions of limiting
canonical momentum $\pm q^0_{\alpha\nu}$ contribute to the independent $-1/2$ holon
$(\alpha\nu =c\nu)$ or $-1/2$ spinon $(\alpha\nu =s\nu)$ spectral function and to the
momentum of the $\alpha\nu = c0,\,s1$ spectral functions associated with the elementary
processes (A), through the FP-scattering-center phase factors.

The operators associated with the pseudofermion spectral functions can be written in the
corresponding direct-product subspaces as,
\begin{eqnarray}
{\tilde{\Theta}}_{k-l\Delta P^F_{\alpha\nu},\,\alpha\nu}^{l,NF,i} & = & \sum_f \langle
f.L;\,\alpha\nu\vert{\tilde{\Theta}}_{k-l\Delta P^F_{\alpha\nu},\,\alpha\nu}^{l,NF,i}
\vert GS\rangle \,\vert f.L;\,\alpha\nu\rangle\langle GS\vert \, ; \hspace{0.25cm}
\alpha\nu= c0,\,s1 \, ,
\nonumber \\
{\tilde{\Theta}}_{k-l\Delta P^0_{\alpha\nu},\,\alpha\nu}^{l,NF,i} & = & \sum_f \langle
f.L;\,\alpha\nu\vert{\tilde{\Theta}}_{k-l\Delta P^0_{\alpha\nu},\,\alpha\nu}^{l,NF,i}
\vert GS\rangle \,\vert f.L;\,\alpha\nu\rangle\langle GS\vert \, ; \hspace{0.25cm}
\alpha\nu\neq c0,\,s1 \, , \nonumber \\
{\tilde{\Theta}}_{\alpha\nu}^{l,F,i}\,\delta_{k,\,l\Delta P^{phF}_{\alpha\nu}} & = &
\sum_f \langle f.L;\,p-h,\alpha\nu\vert{\tilde{\Theta}}_{\alpha\nu}^{l,F,i} \vert
GS\rangle \,\delta_{k,\,l\Delta P^{phF}_{\alpha\nu}}\,\vert
f.L;\,p-h,\alpha\nu\rangle\langle GS\vert \, ; \hspace{0.25cm} \alpha\nu= c0,\,s1 \, .
\nonumber
\end{eqnarray}
Here the $f$ summations run over the states which span such subspaces and the matrix
elements are given by,
\begin{eqnarray}
\langle f.L;\,\alpha\nu\vert{\tilde{\Theta}}_{k-l\Delta
P^F_{\alpha\nu},\,\alpha\nu}^{l,NF,i} \vert GS\rangle & = & \left({1\over
N_a}\right)^{{\cal{N}}_{\alpha\nu}-1\over 2}\,e^{-i\,{\rm sgn} (\Delta
N_{\alpha\nu}^{NF})[\,\sum_{j'=0}^{\vert\Delta
N_{\alpha\nu}^{NF}\vert+N^{phNF}_{\alpha\nu}-1}-\sum_{j'=\vert\Delta
N_{\alpha\nu}^{NF}\vert+N^{phNF}_{\alpha\nu}}^{{\cal{N}}_{\alpha\nu}-1}]
j'a^0_{\alpha\nu}\,{\bar{q}}_{j'}}\nonumber \\
& \times & \delta_{k,\,l[\Delta P^F_{\alpha\nu}+{\rm sgn} (\Delta
N_{\alpha\nu}^{NF})(\sum_{j'=0}^{\Delta
N_{\alpha\nu}^{NF}+N^{phNF}_{\alpha\nu}-1}-\sum_{j'=\Delta
N_{\alpha\nu}^{NF}+N^{phNF}_{\alpha\nu}}^{{\cal{N}}_{\alpha\nu}-1}){\bar{q}}_{j'}]} \, ;
\hspace{0.5cm} \alpha\nu = c0,\,s1 \, ,
\nonumber \\
\langle f.L;\,\alpha\nu\vert{\tilde{\Theta}}_{k-l\Delta
P^0_{\alpha\nu},\,\alpha\nu}^{l,NF,i} \vert GS\rangle & = & \left({1\over
N_a}\right)^{{\cal{N}}_{\alpha\nu}-1\over
2}\,e^{-i\sum_{j'=0}^{{\cal{N}}_{\alpha\nu}-1}j'\,{\bar{q}}_{j'}}\nonumber \\
& \times & \delta_{k,\,l[\Delta P^0_{\alpha\nu}+c_{\alpha\nu}
\sum_{j'=0}^{{\cal{N}}_{\alpha\nu}-1}{\bar{q}}_{j'}]} \, ; \hspace{0.5cm} \alpha\nu\neq
c0,\,s1 \, , \label{matrix-el-0}
\end{eqnarray}
for the $\alpha\nu$ pseudofermion spectral functions and
\begin{equation}
\langle f.L;\,p-h,\alpha\nu\vert{\tilde{\Theta}}_{\alpha\nu}^{l,F,i} \vert GS\rangle
\,\delta_{k,\,l\Delta P^{phF}_{\alpha\nu}} = \langle 0\vert
F_{J-GS,\,\alpha\nu}\,F_{p-h,\,\alpha\nu}\, F_{-GS,\,\alpha\nu}^{\dag}\vert
0\rangle\,\delta_{k,\,l\Delta P^{phF}_{\alpha\nu}} \, ; \hspace{0.5cm} \alpha\nu =
c0,\,s1  \, , \label{matrix-el}
\end{equation}
for the $p-h,\alpha\nu = c0,\,s1$ pseudofermion spectral functions.

The simple form of the matrix elements (\ref{matrix-el-0}) follows from the full overlap
of the generators $F_{J-NF,\,\alpha\nu}$ and $F_{NF,\,\alpha\nu}$ with the operator
${\tilde{\Theta}}_{k,\,\alpha\nu}^{l,NF,i}$ for the $\alpha\nu= c0,\,s1$ and
$\alpha\nu\neq c0,\,s1$ branches, respectively. Such a full overlap also justifies that
the corresponding $\alpha\nu$ spectral functions whose expression is given below have a
non-interacting character. In turn, the evaluation of the matrix element
(\ref{matrix-el}) of the spectral function associated with the $\alpha\nu =c0,\,s1$
pseudofermion elementary processes (B,C) is a more involved problem. For the
$\alpha\nu=c0,\,s1$ branches the phase-factor momentum $l\Delta P^0_{\alpha\nu}=l[\Delta
P^{phF}_{\alpha\nu}+\Delta P^F_{\alpha\nu}]$ involves a term, $l\Delta
P^{phF}_{\alpha\nu}$, which arises from the elementary processes (C). Interestingly, the
dynamics associated with the overlap of the $\alpha\nu =c0,\,s1$ state $\langle
f.L;\,p-h,\alpha\nu\vert=\langle 0\vert F_{J-GS,\,\alpha\nu}\, F_{p-h,\,\alpha\nu}\vert$
with the state ${\tilde{\Theta}}_{\alpha\nu}^{l,F,i}\vert GS\rangle
={\tilde{\Theta}}_{\alpha\nu}^{l,F,i}\,F_{GS,\,\alpha\nu}^{\dag}\vert 0\rangle=
F_{-GS,\,\alpha\nu}^{\dag}\vert 0\rangle$ of the matrix element (\ref{matrix-el}) is not
controlled by the operator ${\tilde{\Theta}}_{k,\,\alpha\nu}^{l,NF,i}$ but rather results
from the different discrete canonical-momentum values of the pseudofermion creation and
annihilation operators involved in the generators of each of these states. (For these
branches the expression of the operator ${\tilde{\Theta}}_{k,\,\alpha\nu}^{l,i}$ does not
include that of the generator $F_{p-h,\,\alpha\nu}^{\dag}$, as mentioned above.) Each
discrete canonical momentum value of the pseudofermion operators involved in the
generators of the former state includes an extra overall canonical-momentum shift
$Q_{\alpha\nu} (q)/L$ relative to those of the latter state. If a $\alpha\nu=c0,\,s1$
pseudofermion or pseudofermion hole is created at the {\it Fermi points} by the
elementary processes (B) and thereafter moved from there by the elementary processes (C)
generated by the operator $F_{p-h,\,\alpha\nu}$, the dynamics associated with the overlap
of the excited-state occupancy configurations generated by the latter processes with the
ground-state generator is controlled by the orthogonality catastrophe that occurs in the
matrix element (\ref{matrix-el}) due to the overall phase shift $Q_{\alpha\nu}(q)/2$.
Such a matrix element involves $N^0_{\alpha\nu} + \Delta N^F_{\alpha\nu}$ $\alpha\nu$
pseudofermions. The occupancy configuration of the state $F_{-GS,\,\alpha\nu}^{\dag}\vert
0\rangle$ corresponds to the densely packed momentum distribution
$N_{\alpha\nu}^{-0}({\bar{q}}_j)$ for $N^0_{\alpha\nu} + \Delta N^F_{\alpha\nu}$
$\alpha\nu$ pseudofermions. The corresponding discrete canonical momentum values
${\bar{q}}_j$ (occupied and unoccupied) are those of the ground state, ${\bar{q}}_j=q_j$.
The occupancy configuration associated with the state $\vert f.L;\,p-h,\alpha\nu\rangle =
F_{p-h,\,\alpha\nu}^{\dag}\,F_{J-GS,\,\alpha\nu}^{\dag}\vert 0\rangle$ also refers to
$N^0_{\alpha\nu} + \Delta N^F_{\alpha\nu}$ $\alpha\nu$ pseudofermions. However, its
discrete canonical momentum values are those of the excited energy eigenstate. This
feature leads to an exotic overlap for the matrix element (\ref{matrix-el}). Such an
overlap is behind the unusual quantum-liquid spectral properties, as further discussed
below and in Ref. \cite{V}. We note that in the absence of the $\alpha\nu = c0,\,s1$
pseudofermion overall phase shifts $Q_{\alpha\nu}(q)/2$, Eq. (\ref{Qcan1j}), the matrix
element (\ref{matrix-el}) would vanish, except for the lowest-peak weight such that
$F_{p-h,\,\alpha\nu}=1$ and thus $\Delta P^{phF}_{\alpha\nu}=0$.

The spectral function expressions are additive in the contributions of each ground-state
- excited-energy-eigenstate transition. For each transition, the available
excited-energy-eigenstate pseudofermion discrete canonical-momentum values are in general
slightly different and given by the functional (\ref{bar-q}). The important point is that
for each ground-state - excited-state transition one knows the precise values of such
discrete pseudofermion canonical momenta. Given these values, the $\alpha\nu$
pseudofermion creation and annihilation operators of the matrix element corresponding to
the specific transition act independently for each $\alpha\nu$ excitation branch. This is
behind the introduction of the above subspace direct product and follows in part from the
factor $\delta_{\alpha\nu,\,\alpha'\nu'}$ on the right-hand side of the pseudofermion
anticommutation relation (\ref{pfacrGS}). Thus, since the pseudofermion creation and
annihilation operators of each $\alpha\nu$ branch act independently for each ground-state
- excited-energy-eigenstate transition, they also do it for the whole spectral function,
which is additive in the contributions of each ground-state - excited-energy-eigenstate
transition. Moreover, as a result of the additive character of the energy in terms of
$\alpha\nu$ pseudofermion, independent $-1/2$ holon, and independent $-1/2$ spinon single
energies and of the corresponding expression of each J-CPHS ensemble subspace as the
direct product of the above considered subspaces, the excited-energy-eigenstate
wave-functions of the ground-state normal-ordered 1D Hubbard model factorize. It follows
that the spectral functions $B^{l,i} (k,\,\omega)$ of Eq. (\ref{ABONj+-cp}), generated by
transitions from the ground state to a given J-CPHS ensemble subspace, can be expressed
as a convolution of pseudofermion spectral functions, one for each branch with finite
occupancy in such a subspace and for the independent $-1/2$ holons and/or independent
$-1/2$ spinons, if they have finite occupancy in the same subspace, and two functions for
the particular case of the $\alpha\nu =c0,\,s1$ branches, as discussed above. It follows
from the form of the matrix elements given in Eq. (\ref{matrix-el}) that the contribution
of the corresponding $p-h,\alpha\nu = c0,\,s1$ pseudofermion spectral functions to the
weight overlaps is more involved than that of the remaining pseudofermion, independent
$-1/2$ holon, and independent $-1/2$ spinon spectral functions.

For each J-CPHS ensemble subspace, we introduce a dimension $D$ associated with the
elementary processes (A),
\begin{equation}
D = \sum_{\alpha\nu}\theta ({\cal{N}}_{\alpha\nu}) \, , \label{Dim}
\end{equation}
where the numbers ${\cal{N}}_{\alpha\nu}$ are defined in Eq. (\ref{Than}). For a general
J-CPHS subspace the function $B^{l,i} (k,\,\omega)$ of Eq. (\ref{ABONj+-cp}) can be
written as a convolution of the $p-h,c0,\,s1$ pseudofermion spectral function, $p-h,s1$
pseudofermion spectral function, one ${\cal{N}}_{\alpha\nu}$-$\alpha\nu$ pseudofermion
spectral function for each of the $D$ branches such that ${\cal{N}}_{\alpha\nu}>0$,
independent $-1/2$ holon spectral function, and independent $-1/2$ spinon spectral
function. Thus, let us provide the general expressions of the spectral functions
corresponding to such a general J-CPHS ensemble subspace.

The $p-h,\alpha\nu = c0,\,s1$ pseudofermion spectral function associated with the
elementary processes (B,C) is given by,
\begin{eqnarray}
B^{l,i}_{Q_{\alpha\nu}} (k,\omega) & = & \sum_{J-CPHS-\alpha\nu-(C)}\vert\langle 0\vert
F_{J-GS,\,\alpha\nu}\,F_{p-h,\,\alpha\nu}\,
F_{-GS,\,\alpha\nu}^{\dag}\vert 0\rangle\vert^2\nonumber \\
& \times & \delta (\omega -l\Delta E^{phF}_{\alpha\nu})\,\delta_{k,\, l\Delta
P^{phF}_{\alpha\nu}} \, ; \hspace{0.5cm} \alpha\nu = c0,\,s1 \hspace{0.25cm} l = \pm 1 \,
, \hspace{0.25cm} i=0,1,2,... \, , \label{Blikom}
\end{eqnarray}
where the energy and momentum spectra are given below and the matrix element is that of
Eq. (\ref{matrix-el}). The summation $\sum_{J-CPHS-\alpha\nu-(C)}$ runs over the J-CPHS
ensemble subspace $\alpha\nu= c0,\,s1$ pseudofermion occupancy configurations generated
by the elementary processes (C). The indices $Q_{c0}$ and $Q_{s1}$ remind us that the
overall phase shifts of Eq. (\ref{Qcan1j}) have a specific value for each ground-state -
excited-energy-eigenstate transition.

In turn, it follows from the form of the matrix elements of Eq. (\ref{matrix-el-0}) that
the $\alpha\nu$ pseudofermion spectral function $B^{l,NF,i}_{\alpha\nu} (k,\omega)$
associated with the elementary processes (A) has a non-interacting character and reads,
\begin{equation}
B^{l,NF,i}_{\alpha\nu} (k,\omega) = \left({1\over
N_a}\right)^{{\cal{N}}_{\alpha\nu}-1}\sum_{J-CPHS-\alpha\nu-(A)}\delta (\omega -l\Delta
E_{\alpha\nu})\,\delta_{k,\, l\Delta P_{\alpha\nu}} \, ; \hspace{0.50cm} l = \pm 1 \, ,
\hspace{0.25cm} i=0,1,2,... \, , \label{B-non-c0s1}
\end{equation}
both for the $\alpha\nu =c0,\,s1$ and $\alpha\nu\neq c0,\,s1$ branches. Here the
summation $\sum_{J-CPHS-\alpha\nu-(A)}$ runs over the J-CPHS ensemble subspace
$\alpha\nu$ pseudofermion occupancy configurations generated by the elementary processes
(A). For the $\alpha\nu\neq c0,\,s1$ branches and densities in the domains $0<n<1$ and
$0<m<n$ the number of the latter occupancy configurations is given by
$D_{\alpha\nu}={N^*_{\alpha\nu}-N_{\alpha\nu}^{F}\choose N_{\alpha\nu}^{NF}}$ and thus
can be written as follows,
\begin{eqnarray}
D_{c\nu} & = & {N_a - N + 2\sum_{\nu'=\nu +1}^{\infty} (\nu' -\nu) N_{c\nu'} +
2L_{c,\,-1/2}-N_{c\nu}^{F}\choose N_{c\nu}^{NF}}  \, ; \hspace{0.5cm} \nu > 0 \, ,
\nonumber \\
D_{s\nu} & = & {N_{\uparrow}-N_{\downarrow} + 2\sum_{\nu'=\nu +1}^{\infty} (\nu' -\nu)
\Delta N_{c\nu'} + 2L_{c,\,-1/2}-N_{s\nu}^{F}\choose N_{s\nu}^{NF}} \, ; \hspace{0.5cm}
\nu > 1 \, , \label{N-choose-N}
\end{eqnarray}
where the values of $N$, $N_{\uparrow}$, and $N_{\downarrow}$ are those of the
corresponding excited-state CPHS ensemble subspace. We recall that for the $c\nu\neq c0$
(and $s\nu\neq s1$) branches and electronic density $n=1$ (and spin density $m=0$) the
spectral function $B^{l,NF,i}_{c\nu} (k,\omega)$ (and $B^{l,NF,i}_{s\nu} (k,\omega)$) of
Eq. (\ref{B-non-c0s1}) does not exist.

Finally, the form of the operator ${\tilde{\Theta}}_{\alpha\nu}^{l,F,i}$ implies that
$\langle GS\vert F_{F,\,\alpha\nu}{\tilde{\Theta}}_{\alpha\nu}^{l,F,i} \vert GS\rangle=1$
for the $\alpha\nu\neq c0,\,s1$ branches with finite occupancy in the J-CPHS ensemble
subspace. This together with the non-interacting character of the $-1/2$ Yang holons and
$-1/2$ HL spinons is behind the form of the independent $-1/2$ holon $(\alpha =c)$ and
independent $-1/2$ spinon $(\alpha =s)$ spectral function, which reads,
\begin{equation}
B^{l,i}_{\alpha,\,-1/2} (k,\omega) = {1\over {\cal {C}}_{\alpha}}\,\delta (\omega
-lE_{\alpha})\,\delta_{k,\,lP_{\alpha}} \, ; \hspace{0.5cm} \alpha = c,\,s \, .
\label{BYHL}
\end{equation}
Here the coefficient ${\cal{C}}_{\alpha}$ is given in Eq. (\ref{Calpha}). While all
spectral functions provided in Eqs. (\ref{B-non-c0s1}) and (\ref{BYHL}) have a
non-interacting character, the $p-h,c0$ and $p-h,s1$ pseudofermion spectral functions of
Eq. (\ref{Blikom}) correspond to a more complex problem. The latter functions are further
studied in Ref. \cite{V} for the metallic phase.

The $\alpha\nu$ pseudofermion energy spectrum $\Delta E_{\alpha\nu}$ on the right-hand
side of Eqs. (\ref{Blikom}) and (\ref{B-non-c0s1}) can be expressed in terms of the
bare-momentum distribution function deviations. The energy spectra $\Delta E_{\alpha\nu}$
and $\Delta E^{phF}_{\alpha\nu}$ appearing in the latter equations and in Eq.
(\ref{Blikom}), respectively, and the the independent $-1/2$ holon $(\alpha =c)$ and
independent $-1/2$ spinon $(\alpha =s)$ energy $E_{\alpha}$ of Eq. (\ref{BYHL}) read,
\begin{eqnarray}
\Delta E_{\alpha\nu} & = & \sum_{q_j=-q_{\alpha\nu}^0}^{+q_{\alpha\nu}^0}\, \Delta
N_{\alpha\nu}^{NF} (q_j)\,\epsilon_{\alpha\nu} (q_j) \, , \nonumber \\
\Delta E^{phF}_{\alpha\nu} & = & {2\pi\over
L}\,v_{\alpha\nu}\,[m_{\alpha\nu,\,+1}+m_{\alpha\nu,\,-1}]\, ;
\hspace{0.5cm} \alpha\nu = c0,\,s1 \, , \nonumber \\
E_{\alpha} & = & \mu_{\alpha}\,[L_{\alpha,\,-1/2}+ \delta_{\alpha,\,c}\,N^F_{c1} +
\sum_{\nu =2}^{\infty}\nu N^F_{\alpha\nu}] \, ; \hspace{0.5cm} \alpha = c,\,s \, ;
\hspace{0.5cm} \mu_c = 2\mu \, , \hspace{0.5cm} \mu_s = 2\mu_0\,H \, . \label{DEan}
\end{eqnarray}
The momentum spectra corresponding to such energy spectra are given by,
\begin{eqnarray}
\Delta P_{\alpha\nu} & = & \sum_{q_j=-q_{\alpha\nu}^0}^{+q_{\alpha\nu}^0}\, \Delta
N_{\alpha\nu}^{NF} (q_j)\,q_j + \Delta P^F_{\alpha\nu} \, ; \hspace{0.50cm} \Delta
P^{phF}_{\alpha\nu} = {2\pi\over L}\,[m_{\alpha\nu,\,+1}-m_{\alpha\nu,\,-1}] \, ;
\hspace{0.25cm} \alpha\nu =c0,\,s1 \, , \nonumber \\
\Delta P^F_{c0} & = & 4k_F\Bigl[\,\Delta J^F_{c0} + \sum_{\nu =1}^{\infty}\,J^I_{c\nu} +
\sum_{\nu =2}^{\infty}\,J^I_{s\nu}\Bigr] \, ; \hspace{0.50cm}  \Delta
P^F_{s1}=2k_{F\downarrow}\Bigl[\,\Delta J^F_{s1} - 2\sum_{\nu
=2}^{\infty}\,J^I_{s\nu}\Bigr] ; \hspace{0.25cm} 0<n<1 , \hspace{0.25cm} 0<m<n
\, , \nonumber \\
\Delta P_{c\nu} & = & \sum_{q_j=-q_{c\nu }^0}^{+q_{c\nu }^0} \,\Delta N_{c\nu}^{NF}
(q_j)\, [(1+\nu)\pi -q_j] \, ; \hspace{0.50cm} \Delta P_{s\nu} =
\sum_{q_j=-q_{\alpha\nu}^0}^{+q_{\alpha\nu}^0}\, \Delta N_{\alpha\nu}^{NF} (q_j)\,q_j \,
; \hspace{0.25cm} \alpha\nu\neq
c0,\,s1 \, , \nonumber \\
P_{c} & = & \pi\,[L_{c,\,-1/2}+\sum_{\nu =1}^{\infty}\nu N^F_{c\nu}] \, ; \hspace{0.50cm}
P_s = 0 \, . \label{noPONE}
\end{eqnarray}
In these equations $\epsilon_{c\nu} (q_j)=2\nu\mu + \epsilon^0_{c\nu} (q_j)$ for $\nu
>0$, $\epsilon_{s\nu} (q_j)=2\nu\mu_0\,H + \epsilon^0_{s\nu} (q_j)$ for $\nu >1$, the
bands $\epsilon_{\alpha\nu} (q_j)$ for $\alpha\nu = c0,\, s1$ and $\epsilon^0_{\alpha\nu}
(q_j)$ for $\alpha\nu\neq c0,\, s1$ are defined by Eqs. (C.15)-(C.21) of Ref. \cite{I},
the small energy $\Delta E^{phF}_{\alpha\nu}$ is such that $m_{\alpha\nu,\,\pm 1}$ is the
number of elementary $\alpha\nu =c0,\,s1$ pseudofermion particle-hole processes (C)
considered above, $v_{\alpha\nu}\equiv v_{\alpha\nu}(q^0_{F\alpha\nu})$, and
$v_{\alpha\nu}(q) =
\partial\epsilon_{\alpha\nu}(q)/\partial q$.

As further discussed in Ref. \cite{V}, for densities $0<n<1$ and $0<m<n$ the elementary
processes (C) leading to the spectral-function singular features include contributions
from small but finite values of $m_{\alpha\nu,\,\pm 1}/N_a$ as $N_a\rightarrow\infty$.
For $n=1$ (and $m=0$) the latter processes do not exist for the $c0$ (and $s1$) branch
and thus $\Delta E^{phF}_{c0}=0$ and $\Delta P^{phF}_{c0}=0$ (and $\Delta E^{phF}_{s1}=0$
and $\Delta P^{phF}_{s1}=0$).

Let us consider the general situation when the J-CPHS ensemble subspace has finite
occupancy for the $c0$ and $s1$ pseudofermion branches, $D-2>0$ $\alpha\nu\neq c0,\,s1$
pseudofermion branches, independent $-1/2$ holons, and independent $-1/2$ spinons. In
this case the functions $B^{l,i} (k,\,\omega)$ of Eq. (\ref{ABONj+-cp}) can be written
as,
\begin{eqnarray}
B^{l,i} (k,\,\omega) & = & {1\over G_C}\,G^{l,i} (k,\,\omega) \nonumber \\
& = & \Bigl(\prod_{\alpha =c,\,s}{1\over
C_{\alpha}}\Bigr)\,\Bigl(\prod_{j=1}^{D}\left({1\over
N_a}\right)^{{\cal{N}}_{\alpha\nu_j}}
\Bigl[\sum_{J-CPHS-\alpha\nu_j-(A)}\Bigr]\Bigr)\,{1\over N_a}
\sum_{k'}\sum_{\omega'}\nonumber \\
& \times & B^{l,i}_{Q_{c0}} \Bigl(k-l\sum_{j=1}^{D}\Delta P_{\alpha\nu_j} -l\sum_{\alpha
=c,\,s}P_{\alpha} -k',\omega -l\sum_{j=1}^{D}\Delta E_{\alpha\nu_j}-l\sum_{\alpha
=c,\,s}E_{\alpha}-
\omega'\Bigr)\nonumber \\
& \times & B^{l,i}_{Q_{s1}} \Bigl(k',\omega'\Bigr) \, ; \hspace{0.5cm} C_c \equiv
{\cal{C}}_c \, ; \hspace{0.5cm} C_s \equiv G_C\,{\cal{C}}_s \, ; \hspace{0.5cm}
i=0,1,2,... \, ; \hspace{0.5cm} l = \pm 1 \, , \label{ABONjlDNLYHL}
\end{eqnarray}
where
\begin{eqnarray}
G^{l,i} (k,\,\omega) & = & \sum_{k_1}\sum_{\omega_1}B^{l,i}_{Q_{c0}} (k-k_1,\omega
-\omega_1)\Bigl[\prod_{j=1}^{D}{1\over N_a}\sum_{k_{j+1}}\sum_{\omega_{j+1}}
B^{l,NF,i}_{\alpha\nu_j} (k_j-k_{j+1},\omega_j
-\omega_{j+1})\Bigr]\nonumber \\
& \times &\Bigl[\prod_{j=D+1}^{D+2} B^{l,i}_{\alpha_{j},\,-1/2} (k_{j}-k_{j+1},\omega_{j}
-\omega_{j+1})\Bigr]\,{1\over N_a} B^{l,i}_{Q_{s1}} (k_{D+3},\omega_{D+3}) \, , \nonumber
\\
G_C & = & \Bigl(\prod_{\alpha
=c,\,s}{\cal{C}}_{\alpha}\Bigr)\Bigl[\sum_J\sum_{k}\int_{0}^{l\infty}d\omega\,G^{l,i}
(k,\,\omega)\Bigl]\Big/\Bigr[\sum_f\sum_{j=1}^{N_a}\,\vert\langle f.L\,;C\vert
{\tilde{\Theta}}_{{{\cal{N}}_i},\,j}^{l}\vert GS\rangle\vert^2\Bigr] \, . \nonumber
\end{eqnarray}
Here the coefficient $G_C$, which also appears in the operator expression of Eq.
(\ref{Than}), has a uniquely defined value for each CPHS ensemble subspace,
${\tilde{\Theta}}_{{{\cal{N}}_i},\,j}^{l}$ is the corresponding operator on the
right-hand side of Eq. (\ref{ONjtil}), the summation $\sum_f$ runs over all energy
eigenstates of the CPHS ensemble subspace, and the summation $\sum_J$ is over all J-CPHS
ensemble subspaces of that subspace. Moreover, the pseudofermion spectral functions
appearing in the $G^{l,i} (k,\,\omega)$ expression are given in Eqs. (\ref{Blikom}) and
(\ref{B-non-c0s1}), the independent $-1/2$ holon and $-1/2$ spinon spectral functions are
provided in Eq. (\ref{BYHL}), $\alpha_{D+1}=c$ and $\alpha_{D+2}=s$ labels the
independent $-1/2$ holons and independent $-1/2$ spinons, respectively, the momenta $k_1,
k_2, ...,k_{D+3}$ and energies $\omega_1, \omega_2, ...,\omega_{D+3}$ correspond to
summation variables, and the index $\alpha\nu_j$, where $j=1,...,D$, is such that
$\alpha\nu_1=c0$, $\alpha\nu_2=s1$, and for $j=3,...,D$ $\alpha\nu_j$ refers to the $D-2$
$\alpha\nu\neq c0,\,s1$ pseudofermion branches such that $N^{NF}_{\alpha\nu}>0$ for the
J-CPHS ensemble subspace. To reach the second expression of Eq. (\ref{ABONjlDNLYHL}) from
the expression for $G^{l,i} (k,\,\omega)/G_C$, we used the non-interacting form of the
spectral functions given in Eqs. (\ref{B-non-c0s1}) and (\ref{BYHL}) to perform $D+2$
momentum and energy summations. It follows that the general spectral function $B^{l,i}
(k,\,\omega)$ of Eq. (\ref{ABONj+-cp}) can be written as a convolution of the $p-h,c0$
and $p-h,s1$ pseudofermion spectral functions alone, as given in the second expression of
Eq. (\ref{ABONjlDNLYHL}). We recall that for the $i=0$ function $B^{l,0} (k,\,\omega)$
the value of the coefficient $G_C$ is independent of $U/t$ and for the dominant CPHS
ensemble subspaces considered in Sec. IV corresponding to that function, it reads $G_C
=1$ for all values of $U/t$.

For J-CPHS subspaces with no finite pseudofermion occupancy for the $\alpha\nu\neq
c0,\,s1$ pseudofermion branches and/or no independent $-1/2$ holon and/or independent
$-1/2$ spinon occupancy, the spectral function $B^{l,i} (k,\,\omega)$ has the same
general form as in Eq. (\ref{ABONjlDNLYHL}), except for the absence in the expression of
$G^{l,i} (k,\,\omega)$ of the spectral functions corresponding to the missing branches
and/or quantum object types. Note also that the expression for $G^{l,i} (k,\,\omega)$ and
thus for $B^{l,i} (k,\,\omega)=G^{l,i} (k,\,\omega)/G_C$ given in the unnumbered equation
after Eq. (\ref{ABONjlDNLYHL}) is valid for electronic density $n=1$ (and spin density
$m=0$), provided that for $j=3,...,D$ the index $\alpha\nu_j$ refers to the $D-2$
$s\nu\neq s1$ (and $c\nu\neq c0$) pseudofermion branches such that $N^{NF}_{\alpha\nu}>0$
for the J-CPHS ensemble subspace. Moreover, for $n=1$ (and $m=0$) one must use $c0$ (and
$s1$) pseudofermion spectral functions specific to the corresponding excitation spectrum.
These functions are studied elsewhere. However, the second expression of Eq.
(\ref{ABONjlDNLYHL}) refers to densities in the domains $0<n<1$ and $0<m<n$ only.

The probability amplitudes $\vert\langle 0\vert
F_{-GS,\,\alpha\nu}\,F_{p-h,\,\alpha\nu}^{\dag}\,F_{J-GS,\,\alpha\nu}^{\dag} \vert
0\rangle\vert^2$ associated with the matrix element (\ref{matrix-el}) which appear in
expression (\ref{Blikom}) have the following general form,
\begin{equation}
\left|\langle 0\vert
F_{-GS,\,\alpha\nu}\,F_{p-h,\,\alpha\nu}^{\dag}\,F_{J-GS,\,\alpha\nu}^{\dag} \vert
0\rangle\right|^2  = \left|\langle 0\vert f_{{{q'}_{N^{0}_{\alpha\nu}+\Delta
N_{\alpha\nu}^F},\,\alpha\nu}}\cdots f_{{{q'}_1},\,\alpha\nu}\,f^{\dag
}_{{{\bar{q}}_1},\,\alpha\nu}\cdots f^{\dag }_{{{\bar{q}}_{{N^{0}_{\alpha\nu}+\Delta
N_{\alpha\nu}^F}}},\,\alpha\nu}\vert 0\rangle \right|^2  \, , \label{me}
\end{equation}
where $\alpha\nu = c0 ,\, s1$. In expression (\ref{me}) we have considered that the
ground state corresponds to pseudofermion annihilation operators and the pseudofermion
operators left for the excited energy eigenstate are of creation character. At this
stage, for the evaluation of $\cal{N}$-electron spectral functions, the main problem
remaining is the computation of the non-trivial probability amplitude (\ref{me}), which
can be expressed by the following determinant,
\begin{equation}
\left|\left|
\begin{array}{llcl} \{f^{\dag }_{{{\bar{q}}_1},\,\alpha\nu}\,
,f_{{{q'}_1},\,\alpha\nu}\} & \{f^{\dag }_{{{\bar{q}}_1},\,\alpha\nu}\,
,f_{{{q'}_2},\,\alpha\nu}\} & \cdots & \{f^{\dag }_{{{\bar{q}}_1},\,\alpha\nu}\,
,f_{{{q'}_{N^{0}_{\alpha\nu}+\Delta N_{\alpha\nu}^F},\,\alpha\nu}}\} \\
\{f^{\dag }_{{{\bar{q}}_2},\,\alpha\nu}\, ,f_{{{q'}_1},\,\alpha\nu}\} & \{f^{\dag
}_{{{\bar{q}}_2},\,\alpha\nu}\, ,f_{{{q'}_2},\,\alpha\nu}\} & \cdots & \{f^{\dag
}_{{{\bar{q}}_2},\,\alpha\nu}\,
,f_{{{q'}_{N^{0}_{\alpha\nu}+\Delta N_{\alpha\nu}^F},\,\alpha\nu}}\} \\
\multicolumn{4}{c}\dotfill\\
\{f^{\dag }_{{{\bar{q}}_{N^{0}_{\alpha\nu}+\Delta N_{\alpha\nu}^F}},\,\alpha\nu}\,
,f_{{{q'}_1},\,\alpha\nu}\} & \{f^{\dag }_{{{\bar{q}}_{N^{0}_{\alpha\nu}+\Delta
N_{\alpha\nu}^F}},\,\alpha\nu}\, ,f_{{{q'}_2},\,\alpha\nu}\} & \cdots & \{f^{\dag
}_{{{\bar{q}}_{N^{0}_{\alpha\nu}+\Delta N_{\alpha\nu}^F}},\,\alpha\nu}\,
,f_{{{q'}_{N^{0}_{\alpha\nu}+\Delta N_{\alpha\nu}^F}},\,\alpha\nu}\}
\end{array} \right| \right|^2  \, , \label{det1}
\end{equation}
for the $\alpha\nu = c0,\,s1$ branches. This result is justified by the following
pseudofermion properties. First, the pseudofermions have no residual-interaction energy
terms, as discussed in Refs. \cite{V,S}. Second, the canonical-momentum shift
$Q_{\alpha\nu} (q_j)/L$, which under the ground-state - excited-energy-eigenstate
transition involves all $\alpha\nu =c0,\,s1$ pseudofermions of the initial ground state,
is a zero-energy process \cite{S}. Third, the elementary pseudofermion processes (C)
correspond to $\alpha\nu =c0,\,s1$ pseudoparticle particle-hole processes whose energy
spectrum is of non-interacting character for the pseudoparticles, what implies that the
energy spectrum $\Delta E^{phF}_{\alpha\nu}=[2\pi/L]\,v_{\alpha\nu}\,m_{\alpha\nu}$ of
Eq. (\ref{DEan}) remains linear in $m_{\alpha\nu}$ for small finite values of
$m_{\alpha\nu}/N_a$ as $N_a\rightarrow\infty$ \cite{V}.

In spite of the non-interacting form of the determinant (\ref{det1}), the unusual
pseudofermion anticommutation relations (\ref{pfacrGS}) give rise to unusual physics, in
the form of an orthogonality catastrophe. (The absence of such an orthogonality
catastrophe would require that $Q_{\alpha\nu} (q)/2=0$.) Indeed, replacement of the
anticommutator (\ref{pfacrGS}) in the determinant (\ref{det1}) leads to,
\begin{eqnarray}
& & \Big({1\over N^*_{\alpha\nu}}\Bigr)^{2[N^{0}_{\alpha\nu}+\Delta
N_{\alpha\nu}^F]}\,\Bigl[ \prod_{j=1}^{N_{\alpha\nu}^*}\,\sin^2 \Big(
{N_{\alpha\nu}^{ph}(q_j)\,[Q_{\alpha\nu}(q_j)-\pi]+\pi\over
2}\Big)\Bigr]\nonumber \\
& \times & \left| \left| \begin{array}{llcl} {1\over\sin\Big({{{\bar{q}}_1}-{{q'}_1}\over
2}\Big)} & {1\over\sin\Big({{{\bar{q}}_1}-{{q'}_2}\over 2}\Big)} & \cdots &
{1\over\sin\Bigl({{{\bar{q}}_1}- {{q'}_{N^{0}_{\alpha\nu}+\Delta
N_{\alpha\nu}^F}}\over 2}\Bigr)} \\
{1\over\sin\Big({{{\bar{q}}_2}-{{q'}_1}\over 2}\Big)} &
{1\over\sin\Big({{{\bar{q}}_2}-{{q'}_2}\over 2}\Big)} & \cdots &
{1\over\sin\Big({{{\bar{q}}_2}- {{q'}_{N^{0}_{\alpha\nu}+\Delta
N_{\alpha\nu}^F}}\over 2}\Big)} \\
\multicolumn{4}{c}\dotfill\\
{1\over\sin\Big({{{\bar{q}}_{N^{0}_{\alpha\nu}+\Delta N_{\alpha\nu}^F}}-{{q'}_1}\over
2}\Big)} & {1\over\sin\Big({{{\bar{q}}_{N^{0}_{\alpha\nu}+\Delta
N_{\alpha\nu}^F}}-{{q'}_2}\over 2}\Big)} & \cdots &
{1\over\sin\Big({{{\bar{q}}_{N^{0}_{\alpha\nu}+\Delta N_{\alpha\nu}^F}}-
{{q'}_{N^{0}_{\alpha\nu}+\Delta N_{\alpha\nu}^F}}\over 2}\Big)}
\end{array} \right| \right|^2 \, ,
\label{det2}
\end{eqnarray}
for the $\alpha\nu = c0,\,s1$ branches, where the overall phase shift $Q_{\alpha\nu}
(q_j)/2$ is given in Eq. (\ref{Qcan1j}) and the bare-momentum distribution function
$N_{\alpha\nu}^{ph}(q_j)$ is such that
$N_{\alpha\nu}^{ph}(q_j)={\cal{N}}_{\alpha\nu}^{ph}({\bar{q}}_j)$. Here
${\cal{N}}_{\alpha\nu}^{ph}({\bar{q}}_j)$ is the pseudofermion canonical-momentum
distribution function given in Eq. (\ref{NanphTh}), which includes the low-energy and
small-momentum $\alpha\nu = c0,\,s1$ pseudofermion particle-hole processes (C). The
determinant of Eq. (\ref{det2}) can be rewritten as,
\begin{eqnarray}
& & \Big({1\over N^*_{\alpha\nu}}\Bigr)^{2[N^{0}_{\alpha\nu}+\Delta N_{\alpha\nu}^F]}\,
\prod_{j=1}^{N_{\alpha\nu}^*}\,\sin^2 \Big(
{N_{\alpha\nu}^{ph}(q_j)\,[Q_{\alpha\nu}(q_j)-\pi]
+\pi\over 2}\Big) \nonumber \\
& \times & \prod_{j=1}^{N_{\alpha\nu}^*}\prod_{i=1}^{N_{\alpha\nu}^*}\,\theta (i-j)\,
\sin^2\Bigl({N_{\alpha\nu}^{-0}({q'}_j) N_{\alpha\nu}^{-0}({q'}_i)[{q'}_j - {q'}_i -\pi]
+ \pi\over 2}\Bigr) \nonumber \\
& \times & \prod_{j=1}^{N_{\alpha\nu}^*}\prod_{i=1}^{N_{\alpha\nu}^*}\,\theta (i-j)\,
\sin^2\Bigl({N_{\alpha\nu}^{ph}({q}_j) N_{\alpha\nu}^{ph}({q}_i)[{\bar{q}}_j -
{\bar{q}}_i -\pi] + \pi\over
2}\Bigr) \nonumber \\
& \times & \prod_{j=1}^{N_{\alpha\nu}^*}\prod_{i=1}^{N_{\alpha\nu}^*}\,{1\over
\sin^{2}\Bigl({N_{\alpha\nu}^{ph}({q}_i) N_{\alpha\nu}^{-0}({q'}_j)[{\bar{q}}_i - {q'}_j
-\pi] + \pi\over 2}\Bigr)} \, ; \hspace{0.5cm} \alpha\nu = c0,\,s1 \, , \label{det3}
\end{eqnarray}
where $N_{\alpha\nu}^{-0}(q_j)= {\cal{N}}_{\alpha\nu}^{-0}({\bar{q}}_j)$ is a densely
packed bare-momentum distribution function whose {\it Fermi points} are given by
$q_{F\alpha\nu,\,\pm 1}=\pm q_{F\alpha\nu}^0\pm [2\pi/L]\Delta N^{0,F}_{\alpha\nu ,\,\pm
1}$ and the corresponding canonical-momentum distribution function
${\cal{N}}_{\alpha\nu}^{-0}({\bar{q}}_j)$ is that of Eq. (\ref{NanphTh}) whose {\it Fermi
points} read ${\bar{q}}_{F\alpha\nu,\,\pm 1}=\pm q_{F\alpha\nu}^0\pm [2\pi/L][\Delta
N^F_{\alpha\nu ,\,\pm 1}\pm Q^{\Phi}_{\alpha\nu} (\pm q_{F\alpha\nu}^0)/2\pi]$. The
expressions (\ref{det1})-(\ref{det3}) are used in Ref. \cite{V} in the derivation of
finite-energy spectral-weight distributions for the model metallic phase.

\section{DISCUSSION AND CONCLUDING REMARKS}
\label{SecVI}

The main result of this paper is the general spectral function expression defined by Eqs.
(\ref{ABONjl-CPHS}) and (\ref{ABONjlDNLYHL}). The expression given in the latter equation
involves the $p-h,c0$ and $p-h,s1$ pseudofermion spectral functions provided in Eq.
(\ref{Blikom}), whose probability amplitude $\vert\langle 0\vert
F_{J-GS,\,\alpha\nu}\,F_{p-h,\,\alpha\nu}\, F_{-GS,\,\alpha\nu}^{\dag}\vert
0\rangle\vert^2$ can be expressed in terms of the determinants of Eqs.
(\ref{det1})-(\ref{det3}). An important aspect of the pseudofermion dynamical theory
introduced in this paper and further developed in Ref. \cite{V} for the metallic phase,
is the different origin of the dynamics associated with the matrix-element overlaps of
the $\alpha\nu=c0,\,s1$ pseudofermion occupancy configurations in the vicinity and away
of the {\it Fermi points}.

In reference \cite{spectral1} the finite-energy spectral function expressions derived  by
use of the pseudofermion dynamical theory introduced here are applied to the study of the
spectral-weight features observed in the quasi-1D organic compound TTF-TCNQ.
Interestingly, one finds quantitative agreement with the observed spectral features for
the whole experimental energy band width. The microscopic mechanisms found in Ref.
\cite{super} by use of our theory are also consistent with the phase diagram observed in
the (TMTTF)$_2$X and (TMTSF)$_2$X series of organic compounds and explain the absence of
superconducting phases in TTF-TCNQ. Our theory is also of interest for the understanding
of the spectral properties of the new quantum systems described by ultra-cold fermionic
atoms on an optical lattice \cite{Zoller}.

\begin{acknowledgments}
We thank Daniel Bozi, Ralph Claessen, Patrick A. Lee, Gerardo Ortiz, and Pedro D.
Sacramento for stimulating discussions and the support of the ESF Science Programme
INSTANS 2005-2010. J.M.P.C. thanks the hospitality and support of MIT and the financial
support of the Gulbenkian Foundation, Fulbright Commission, and FCT under the grant
POCTI/FIS/58133/2004 and K.P. thanks the support of the OTKA grant T049607.
\end{acknowledgments}

\end{document}